\documentclass[12pt,preprint]{aastex}

\received{2001 October 13}
\begin{document}

\newcommand{\coto}{$^{12}$CO(2--1) \/}
\newcommand{\cooz}{$^{12}$CO(1--0) \/}
\newcommand{\thcooz}{$^{13}$CO(1--0) \/}
\newcommand{\thco}{$^{13}$CO \/}
\newcommand{\twco}{$^{12}$CO \/}
\newcommand{\vlsr}{$V_{LSR}$ \/}
\newcommand{\kmss}{km~s$^{-1}$ \/}
\newcommand{\kms}{km~s$^{-1}$}

\title{Bow shocks, wiggling jets, and wide-angle winds: a high resolution study
of the entrainment mechanism of the PV Ceph molecular (CO) outflow}
\author{H\'ector G. Arce\footnote{Now at: California Institute of Technology,
Astronomy Department, MS 105-24, Pasadena, CA 91125. 
\newline New e-mail: harce@astro.caltech.edu} \/
 \& Alyssa A. Goodman}
\affil{Harvard--Smithsonian Center for Astrophysics, 60 Garden St.,
Cambridge, MA 02138}
\email{harce@cfa.harvard.edu, agoodman@cfa.harvard.edu}

\begin{abstract}

We present a new set of high-resolution molecular line maps
of the gas immediately
surrounding various Herbig-Haro (HH) knots of the
giant HH flow HH~315, from the young star PV Cephei.
The observations, aimed at studying the entrainment
 mechanism of the 2.6~pc-long HH~315 flow,
 include IRAM~30~m maps of the $^{12}$CO(2--1), $^{12}$CO(1--0), and
\thcooz lines, with beam sizes
of 11\arcsec, 21\arcsec, and 22\arcsec, respectively.
We compare the morphology and the kinematics of the outflow gas, 
as well as the temperature
and momentum distribution of the molecular outflow with those 
predicted by different entrainment
models. With our detailed study we are able to conclude that 
jet bow shock entrainment 
by an episodic stellar wind, with a time-varying axis,  
produces most of the high-velocity molecular outflow observed far 
from the source.
In addition, near PV Cephei we find evidence for  a
poorly collimated, wide-angle,  molecular outflow {\it and} 
a collimated wiggling jet-like molecular outflow. We propose
that the poorly collimated component is entrained by a wide-angle wind, 
and the collimated component is
entrained by a variable jet with internal working surfaces.
If this picture is true, then
a stellar wind model which allows for the coexistence
of a wide-angle component {\it and} a collimated (jet-like)
stellar wind component is needed to explain the observed
properties of the PV Ceph outflow.
The wiggling axis of the redshifted molecular outflow 
lobe indicates that the outflow ejection axis is changing over time.
We find that the time-scale of the  axis variation
shown by the molecular outflow lobe is about a factor of 10
less than that shown by the large-scale optical HH knots.
\end{abstract}

\keywords{ISM: jets and outflows --- ISM: Herbig-Haro objects ---
ISM: individual(PV Ceph, HH 315) --- stars: formation}

\section{Introduction}

In order to
better understand the effects of  winds from young stars
on the ambient molecular cloud, it
is essential to study how the
wind interacts with its surrounding medium. The best
 evidence that winds from young stars interact
with their surrounding  ambient gas  is the existence of molecular outflows.
Many molecular outflows have masses that are
larger than their powering young star, by a factor of a few up 
to an order of magnitude (e.g., Wu, Huang, \& He~1996). 
It is, therefore, 
highly unlikely that the mass in CO outflows comes directly 
from the forming star and/or the circumstellar  
disk, and so molecular outflows 
consist mainly of swept-up ambient material.

The mechanism by
 which a wind from a young stellar object entrains and accelerates
the ambient gas, thereby producing a molecular outflow, is still a 
matter of debate, 
even though several models have been proposed.
The three most popular entrainment models are the turbulent jet model, 
the bow shock model, and the
wide-angle wind model. 

 In the turbulent jet entrainment model (e.g., Cant\'o \& Raga 1991; 
Stahler 1994; Lizano \& Giovanardi 1995), the ambient gas is 
entrained on the sides of a  jet (or wind) 
through a turbulent viscous mixing layer 
formed by Kelvin-Helmholtz instabilities.

In the bow shock model (e.g., Raga \& Cabrit 1993; Stone \& Norman 1993a; 
Masson \& Chernin 1993;
Stone \& Norman 1994; Suttner et al.~1997; Zhang \& Zheng 1997; Smith, 
Suttner, \& Yorke 1997; Downes
\& Ray 1999; Lee et al.~2001) 
a highly-supersonic  collimated wind or jet propagates into the ambient
medium, forming a bow shock surface at the head of the jet. The jet  
carves into the cloud, and the 
bow shock will
move away from the star, interacting with the ambient gas,  thereby producing 
a molecular outflow
around the jet. Velocity variations in the jet  produce  
bow shocks  along the body of the jet, also referred to as internal working 
surfaces (e.g., 
Raga et al.~1990; Raga \& Kofman 1992; Cant\'o, Raga, \& D'Alessio 2000), 
which can also help entrain
ambient gas.

In the wide-angle wind model (Shu et al.~1991; Li \& Shu 1996; 
Matzner \& McKee 1999; Lee et al. 2000; 2001), 
the outflow is produced when a momentum-conserving wide-angle wind from 
a young stellar object
 interacts with the ambient gas. The wind power is
dependent on polar angle, has a constant velocity and blows into a 
core with radial density profile and angular
dependence.
 The ambient medium is swept up into a shell 
by a shock at the wind bubble-ambient cloud interface. The molecular 
outflow is then identified
as the molecular gas in the entrained shell.

In Figure~1 we summarize the 
expected molecular outflow characteristics  
(i.e., gas morphology, temperature, velocity and momentum distribution), 
for these three  entrainment models.

On the observational front, recent high angular resolution studies 
of molecular outflows
(e.g., Richer, Hills, 
\& Padman 1992; Bence, Richer, \& Padman~1996; Cernicharo \& Reipurth 1996;
Lada \& Fich 1996; Davis, Smith, \& Moriarty-Schieven 1998; 
Shepherd et al.~1998; Gueth \& Guilloteau 1999; 
Yu, Billawala, \& Bally~1999; Davis et al.~2000; Lee et al.~2000; 
Arce \& Goodman 2001b) 
have provided important information on the physical parameters of 
outflows and the entrainment mechanism.

In Arce \& Goodman (2002, hereafter Paper I)
 we  studied the interaction of the
 HH~315 flow with its parent
cloud on large scales (see also Paper I for a brief description of HH~315).
In this paper, we zoom in on the gas immediately surrounding several 
HH knots and the outflow source (PV Ceph), in order to study the 
entrainment mechanism of
the HH~315 outflow. 
In particular,
 we study the temperature distribution, kinematics, momentum distribution,
and morphology of the outflow gas.
The results are then used to compare all of
these observed outflow characteristics with those expected 
from different entrainment models.
Along  with the  information from 
earlier high resolution studies, our results
provide important constraints 
to be considered by future theoretical entrainment models.

In the following section we will describe the observations. 
This is  followed by a section where we
describe our  results. We later use our results to
compare them with  the results of the three  
molecular outflows entrainment models shown in Figure~1. 
Subsequently, we discuss the episodic and
wandering (or wiggling) nature of the HH~315 outflow. 
Lastly, we give a summary of our findings.

\section{Observations}

In order to study in detail the interaction between the HH~315 flow 
and the surrounding gas,
we made high spatial- and velocity- resolution observations of the gas 
around several of the
HH knots in the flow. The data were obtained using the IRAM 30~m telescope 
in Pico Veleta, Spain, in September 1999.
The $^{12}$CO(1--0), $^{12}$CO(2--1), and $^{13}$CO(1--0), 
 lines were observed simultaneously using three spectral line SIS
receivers.  The spectrometer used
was an autocorrelator split in three parts, each connected to a different 
receiver.
The $J= 1 \rightarrow 0$ lines were observed with a spectral resolution of 
40~kHz and 
a band width of 20~MHz, and the
\coto line was observed with a spectral resolution of 80~kHz and
a band width of 40~MHz. The telescope beamwidths (FWHM) at 
$^{12}$CO(1--0), $^{12}$CO(2--1), and $^{13}$CO(1--0) are about
21\arcsec, 11\arcsec, and 22\arcsec, respectively.
The forward efficiency ($F_{eff}$) and main beam efficiency ($B_{eff}$) of the 
 $J= 1 \rightarrow 0$ lines are approximately 0.90 and 0.54, respectively,
and for the \coto \/ line $F_{eff} \sim 0.86$ and $B_{eff} \sim 0.42$
(Wild 1999). Unless it is stated otherwise, 
the intensity scale of the spectral data from
the  IRAM 30~m is in units of main beam temperature ($T_{mb}$), where 
$T_{mb} = (F_{eff}/B_{eff})T_{A}^{*}$ (Wild 1999; Rohlfs \& Wilson 2000).

The on-the-fly mapping technique was used to map three regions of interests 
The telescope in OTF mode
moved across the source at a constant speed of 2\arcsec \/ s$^{-1}$,
while a spectrum was acquired every 2 seconds. 
Table~1 lists 
the center position and the size of each of the three major regions 
(see also Figure~2).
The regions were  all scanned in both the right ascension and declination
 directions.
The separation, in the direction
perpendicular to the scanning direction, between subsequent rows was 4\arcsec.

The telescope was pointed to an OFF position, located at 
R.A. $20^h45^m30.4^s$, decl. 67\arcdeg55\arcmin46.7\arcsec \/ (B1950), 
after every other row, where it 
would observe the OFF position for 10 to 15 sec. 
A temperature calibration was done every 5 to 10 minutes.
Deep observations of the OFF position showed no significant emission.
The different regions were observed several times to improve the 
signal-to-noise in the spectra.  
The raw OTF data were reduced using CLASS.
A baseline was fitted to and subtracted from each spectrum.
The spectral maps of each region were convolved with
 Gaussian beams of different sizes, and
Table~2 shows 
the resultant RMS noise for each spectral line map
convolved with different beams.

\section{Results}

\subsection{The region surrounding HH 315B and HH 315C}

Large scale CO maps of the outflow associated with PV Ceph
(Levreault 1984; Paper I) show that 
the region surrounding the optical knots HH~315B and HH~315C
(region hh315b+c) is the area where most of the emission from the
northern (mostly blueshifted) outflow lobe resides (see Figure~2).
In this section,  we present  high spatial resolution  maps
of hh315b+c, which show   
the structure of the outflowing gas in detail, and its 
spatial relation with the HH knots.

\subsubsection{\coto emission}

Figure~3 shows  \coto velocity-integrated  maps 
of the hh315b+c region, observed with the
IRAM~30~m telescope, for  10 different velocity ranges.
The emission with the most blueshifted velocities is 
shown in Figure~3{\it b} and
the emission with the most redshifted velocities is
shown in Figure~3{\it l}. 
 Notice that not all velocity integration ranges
have the same width. The different
ranges of integration were chosen to group different 
channels with similar \coto emission
structure. The grey objects shown in each velocity map panel 
represent the position of the
different S[II] knots in the region, as presented by 
G\'omez, Kenyon, \& Whitney (1997, hereafter GKW).

The highest-velocity (most blueshifted)
 CO outflow emission
is shown in Figure~3{\it b}. The gas at these velocities
is concentrated at the position of the HH~315B optical knot. 
The lowest (plotted)
contour of integrated emission shows wing-like extensions which
 point towards the position
of the outflow source (PV Ceph). The shape of the contours at these
high velocities is suggestive of a bow shock morphology.

At slightly lower (less blueshifted) velocities, the outflow gas is 
concentrated 
along the main HH~315B optical knot (see Figures~3{\it c} and 3{\it d}).   
The gas emission surrounding HH~315B in Figures~3{\it c} and 
3{\it d} is mainly
concentrated south of the emission peak in Figure~3{\it b}, 
and has an elongated
morphology towards the position of PV Ceph.
This is consistent with a picture where the highest velocities are at the
head of the shock (Figure~3{\it b}), 
where the gas is currently being entrained, and the slower wake of
previously entrained gas is found behind the head of the shock 
(emission in  Figures~3{\it c} and 3{\it d} near HH~315B). 

The most striking feature of the velocity maps shown in Figure~3 
is the bow shock- (or shell-)
like structure of the \coto emission, coincident with the 
bow shock-like HH~315C optical knot.
The head of the CO shell coincides with the region of brightest 
 optical emission from  
HH~315C, and has wings, on both sides of the head, which
point in the general direction of PV Ceph. Hereafter we will refer to this
structure as the HH~315C CO shell. 
The HH 315C CO shell is observed at 
different outflow velocities (see Figure~3), and at the slowest (less
blueshifted) outflow velocities it has a width of about 0.4~pc 
(see Figures 3{\it h}, and 3{\it i}).
The positional coincidence of HH~315C and the outflowing
CO shell, in addition to the bow shock structure
of both the optical knot and the CO outflow, are  highly suggestive that these 
are two different outflow manifestations from the same ejection episode.

It is interesting to note
that even though HH~315C is expected\footnote{Most, but not all, 
HH objects have been found to have velocities between 100 to 200~\kmss
(Hartigan et al.~2000; Reipurth \& Bally 2001). }
 to have a velocity of 
$\sim 100$~\kms, 
 the outflowing gas associated with the HH~315C CO shell has radial
outflow velocities of no more than $\sim 5$~\kmss (which at most could 
indicate maximum true space outflow velocities of about 30~\kms, if we assume 
that the angle between the flow axis and the plane of the sky is about 
$10\arcdeg$). The disparity between the  velocity of optical HH knots
and the velocity of
their corresponding CO outflow has always been a puzzle.
In a momentum-conserving interaction one would expect the more massive
component to end up with a slower velocity than the less massive component.
So, the apparent disparity between optical and molecular outflow velocities
could be explained if the optical flow is much less massive than the molecular
outflow (which is what is expected).
In addition, it could be that the HH object originally 
had a larger velocity  
(the usual 100 to 200~\kms), but it has decelerated as it interacts with the
ambient medium to velocities closer to 30-50~\kmss
---much like the giant HH flow HH~34 (Devine  et al.~1997; 
Cabrit \& Raga 2000). If such is the case, then the apparent disparity  
between optical and CO outflow velocity would be much less (almost negligible).
We will not be sure of the exact kinematics of HH~315C until
spectra and proper motions are measured for this optical knot.
Independent of the kinematics of the HH~315C optical knot,
what is really important to the discussion of this paper is that
the morphology of the outflowing molecular gas in the 
hh315b+c region, along with other factors discussed below, 
strongly suggest that the molecular outflow
is bow shock-driven.

In Figure~3{\it j} we show the  \coto integrated over the velocities between 
1.01 and 1.89~\kms. The ambient cloud  velocity in  this region is
1.5~\kmss (see Paper I),
 and so most of the ``ambient'' cloud is found at this velocity range.
In Figure~3{\it k} we plot the \coto integrated over the velocities between 
1.89 and 2.55~\kms. This velocity range also includes emission from 
``ambient'' cloud
gas, as the gas emission shows an extended, cloud-like, morphology. 
In both Figures~3{\it j} and 3{\it k} the density gradient of the cloud is
clearly seen,  the cloud's edge is in the east and the density increases 
towards the west.
At these ``ambient''  velocities,  there is an increase in the column density
 at the position of the different small knots which comprise the 
HH~315C knot. In addition,  the \coto  contours near the HH~315C knots 
curve, following the bow-like configuration of the optical knots. 
Thus, we see that the HH~315C CO shell structure is also detectable at ambient
cloud velocities ---evidence that the HH~315 flow has altered the ambient 
gas distribution
at distances as far as $\sim 1.5$~pc from the source (see Paper I for more on
this).

The \coto map integrated over the
range of $2.55<v<3.65$~\kmss (Figure~3{\it l}), which is
redshifted with respect to the ambient cloud velocity of the region,
shows a ``clump" of emission coincident with the HH~315C optical knot.
Similar to  what is seen in the other 
(blueshifted) velocity-integrated maps in Figure~3,
the \coto intensity contours  curve, following the
bow-like arrangement of  the HH~315C optical knots.
This  redshifted emission
is presumably outflowing gas 
accelerated by the  ``back side'' of the shock front 
associated with HH~315C.

In Figure~4 we 
show several sample \coto spectra from the hh315b+c region. It is
interesting to note the drastic differences between spectra from regions 
very close to each other.
Also, notice that the outflow emission is not a smooth low-level 
wing at blueshifted 
velocities. Instead,  
it is made of  different  velocity (spectral) components, 
which sometimes are stronger than
the ambient cloud component.

\subsubsection{\thco emission}

The \thco emission in the HH315b+c region has
less complicated velocity structure than the \twco emission.
Similar to what is observed in the large-scale map (see Figure~3 in Paper I), 
the  \thcooz emission at velocities greater than 
1.5~\kmss in this region is very weak ($T_{mb} \lesssim 0.6$~K), 
and hence most of the \thco
emission in this region is  blueshifted compared 
to the cloud's ambient velocity. 
In Figure~5 we show three different \thco velocity-integrated intensity maps. 
The different ranges of integration were chosen to group different channels 
with similar \thcooz emission structure.

The \thco vaguely follows
the \twco bow shock structure associated with HH~315C. 
Both the north and the south bow
shock wings are clearly present in \thco emission, 
yet there is very little \thco emission at the
location of the bow shock apex. The north and south \thco bow wings are 
spatially coincident with
the \twco bow wings with similar velocities.
We do not detect any \thco emission associated with 
the high velocity gas surrounding the
HH~315B knot.  The structure of the high spatial resolution \thco is consistent
with our original hypothesis (Paper I) that the blueshifted northern lobe
of the HH~315 molecular outflow has ``pushed'' aside gas, creating a
shell-like structure at its edges, which is dense enough 
($n \gtrsim 500$~cm$^{-3}$)  that  is detected in $^{13}$CO(1-0) emission.

\subsubsection{Mass}
In Table~3 we list the mass for each of the velocity ranges shown in 
Figure~3. 
The mass was obtained using our high-resolution (IRAM~30~m) \cooz and \thcooz
data, with the procedure described in Paper I. 
The only difference in the procedure is the use of the average excitation
 temperature
($\bar{T}_{ex}$), given in Table~3, to estimate the 
outflow mass of a given 
velocity range (rather than using $T_{ex}=10.5$ for all velocities).
Notice that the mass of the ``ambient cloud'' velocity range 
($1.01 < v < 2.55$~\kms) is approximately the same as the sum of 
the rest of the velocity ranges. This implies that in this 
region the outflow mass is approximately the ``ambient cloud'' mass. 
We first noticed this with our large-scale CO maps (Paper I), and  
its importance is discussed there.

\subsection{The region surrounding HH 215 and PV Ceph}

The other region studied at high resolution with the IRAM~30~m 
telescope is the area
surrounding the outflow source, PV Ceph. This area is 
where most of the southern (redshifted)
outflow resides (see Figure~2). In addition, 
the area observed 
also covers the area north of PV Ceph where 
the chain of five optical knots that make up  HH~215 have been detected (GKW).
We call this whole area the hh215 region.
We note that in this region, the central ambient cloud velocity is 
$v_{LSR, south}=2.5$~\kmss (see Paper I), and so 
outflow velocities in the hh215 region are defined as the observed velocity 
minus 2.5~\kms.

\subsubsection{\coto emission}

In Figure~6
we show the integrated intensity, over four 
different velocity ranges, of the \coto 
in the hh215 region.
Notice that not all velocity integration ranges
have the same width. The different
ranges of integration were chosen to group different velocity 
(spectral) channels with similar \coto emission
structure.

The most blueshifted velocity range in which we detect \coto emission in 
this region
is shown in Figure~6{\it a}. The blueshifted molecular outflow emission in 
this velocity
range  ($-0.15<v<0.74$~\kms) is concentrated 
at the source and north of it. 
The emission detected south (and southeast)
of the
 source in Figure~6{\it a} is from another cloud in the same line of sight
(cloud X) which we detect in our large scale \twco and \thco maps
(see Figures~2 and 3 in Paper I). 
The blueshifted \coto outflow emission in Figure~6{\it a} is very poorly 
collimated
---unlike the outflowing redshifted emission discussed below--- 
nor does it show the nice bow shock
structure observed in the hh315b+c region.

In Figure~6{\it b} we show the (blueshifted) \coto integrated  
emission over the
velocity range between $1.62<v<2.06$~\kms. We do not include an integrated 
velocity map
for velocities between 0.74 and 1.62~\kms, since at these velocities the \twco 
is dominated by
the emission from cloud X all through the mapped region, and so no outflow 
features   
are observed.
The velocity range in Figure~6{\it b} includes velocities as close
 as $\sim 0.5$~\kmss to
the  central ambient velocity of 2.5~\kms, so there is some ambient 
cloud emission
contribution to the \coto emission at these velocities. The contours 
were chosen to show
the brightest features.
It can be seen that the emission peaks near the position of PV Ceph
and extends north with a fan-like structure 
(with an opening angle of $\sim 90\arcdeg$). 
This wide-angle 
structure does not resemble any cloud structure observed in our large
scale maps. Thus, we believe
that the emission north of PV Ceph in Figure~6{\it b} comes from the slowest, 
detectable
(blueshifted) outflow emission in the hh215 region. This slow outflow gas has 
a morphology
which is consistent with it being entrained by a wide-angle wind. 
We further discuss this in \S 4.3.3.

The emission south of PV Ceph in Figure~6{\it b} is ambient cloud emission, 
which we
easily  identify with a structure seen in our large scale maps. Notice 
that this emission south of PV
Ceph in Figure~6{\it b} delineates the walls of the cavity cleared by the 
redshifted
outflow lobe  (Figures~6{\it c} and 6{\it d}).
 This cavity is also observed in
our large-scale map and we discuss its importance in Paper I.

Between velocities 2.06 and 3.16~\kms, the detected \coto 
comes from (extremely optically thick)
ambient emission, which shows  no (or very little) structure.
The slowest detectable redshifted outflow emission is 
at $v \sim 3.2$~\kms, and it is easily identified as outflow gas
 from the obvious non-cloud-like morphology of
the gas emission.

The redshifted outflow gas shows a very peculiar structure, 
very different from the
wide-angle blueshifted outflow emission discussed above.
At the lowest redshifted velocities (Figure~6{\it c}) 
the \coto integrated intensity peaks at
the position of PV Ceph and has a cometary-like extension towards the east, 
and a 
collimated (jet-like) extension towards the south.
The east-west cometary-like structure is only clearly detected 
at the velocities between 
$\sim$ 3.2 and 5.2~\kmss and extends further than the eastern limits of
our map. We believe that this cometary-like structure is associated with
the motion of PV Ceph through the cloud (see Goodman \& Arce 2002).
The north-south structure seen in Figure~6{\it c} 
extends about 1.75\arcmin \/ ($\sim 0.25$~pc),
and resembles a curving (or wiggling) CO jet, 
with an eastward bent  at $\sim 1$\arcmin \/ south of
PV Ceph.

At high (redshifted) outflow velocities 
(see Figure~6{\it d}) we detect a  
north-south \twco structure similar to the one seen in Figure~6{\it c}. 
About 50 to 60\arcsec \/ south of PV Ceph  
there is a ``bump'',  where the 
collimated north-south structure widens. 
As discussed later (\S 4.3.2), 
this bump is most probably due to the entrainment of the ambient gas by
a redshifted counter-episode of HH~215.
South of the bump the jet-like structure
continues, and it ends
about 2\arcmin \/ ($\sim 0.29$~pc) south of PV Ceph, slightly further 
south than the north-south
structure in  Figure~6{\it c}.

\subsubsection{\thco emission}

In Figure~7 we show three different velocity-integrated intensity maps 
of the \thcooz
 emission in the hh215 area. Similar to the other velocity-integrated maps 
presented here,
 the ranges of integration  were chosen to group different velocity 
(spectral) channels with similar \thcooz emission structure.

Figure~7{\it a}
 shows the most blueshifted \thco emission in the hh215 area, at velocities
between 1.73 and 2.28~\kms. 
The velocity range is very narrow (only 0.55~\kmss wide) and the velocities 
are close
to the central ambient gas velocity (2.5~\kms). 
The blueshifted \thco integrated intensity map 
(Figure~7{\it a})
 shows a clear  V-shaped structure with its apex at the position of PV Ceph. 
As shown in Figure~8, this
 \thco structure is coincident with the the optical reflection nebula
north of PV Ceph 
(Cohen et al.~1981; Gledhill, Warren-Smith, \& Scarrot 1987;
Levreault \& Opal 1987; Neckel et al.~1987
Scarrot, Rolph, \& Tadhunter 1991; RBD; GKW), and 
encloses the wide-angle \coto blueshifted outflow seen in Figure~6{\it b}.
Thus, it appears that \thco  in this velocity range
traces the limb-brightened
walls of a wind-blown cavity. We will  discuss  this further in \S 4.3.3.

The other two velocity-integrated intensity maps 
of \thcooz (Figures~7b and 7c) come from redshifted velocities
(compared to the ambient cloud velocity of 2.5~\kms). 
The \thco integrated intensity emission in Figure~7{\it b} 
resembles (and coincides with)
 the east-west cometary-like structure in the \coto integrated intensity
map in Figure~6{\it c}. 
In addition to the east-west structure there is a low-emission 
north-south elongation, which partly coincides with the north-south \coto 
jet-like structure in Figure~6{\it c}, and extends only 
$\sim1.3\arcmin$ \/ (0.18~pc) south of PV Ceph.
Figure~7{\it c} shows the most redshifted \thco emission in the hh215 region.
At these velocities the \thco emission is very weak compared to 
the other velocities
shown, and the maximum emission is just south of PV Ceph.

In Table~4 we list the mass for each of the velocity ranges shown in
Figure~6. 
The mass was obtained using our IRAM~30~m \cooz and \thcooz
data, using the procedure described in Paper I (with $T_{ex}=10.5$ 
for all velocity ranges). 
In Figure~9 we show  sample \coto and \thcooz spectra from the hh215 region.

\subsection{The region surrounding HH 315E}

The sensitivity of our observations in the hh315e region is similar to 
that of the other regions
(see Table~2).
So, if the column density of the outflowing CO gas in the hh315e region 
were similar to that of the other
regions,  then outflowing CO should be easily detectable.  
But, even though HH~315E is the counter-knot of HH~315B 
---where we detect CO with the highest blueshifted outflow velocities--- 
we do not detect  
any outflow emission in the region surrounding the optical knot HH~315E.

In Figure~10 we show an average \coto spectrum of the  hh315e region. 
The spectrum does not show any redshifted low-level wing,
or another velocity component  redshifted from the ambient emission, 
as might be  expected 
for a redshifted CO outflow lobe spectrum.
 Instead, the spectrum shows mainly ambient gas emission from the 
cloud associated with PV Ceph
 (the ``PV Ceph cloud''), which peaks at $v\sim 2.5$~\kmss (see Paper I).
The  bump in the spectrum seen at blueshifted velocities
is due to ``contaminating'' emission from another cloud 
on the same line of sight.
We fit a double Gaussian to the average spectrum and from the fit we obtain 
that the PV Ceph cloud
component (centered at $\sim 2.5$~\kms) has a width (FWHM) of $\sim 0.8$~\kms.
Even if we fit a single Gaussian, the resultant velocity width is 1.1~\kms.

A \coto spectrum with such
a narrow velocity is usually observed in quiescent ambient gas clouds, 
rather than in
 regions affected by stellar outflows.  HH knots are shocks
arising from the interaction
of a high-velocity flow of gas ejected by a young stellar object
and the ambient medium.
Thus, we know  
 that the outflow mass ejection responsible for
the HH~315E optical knot is interacting with its surrounding medium 
because HH~315E is detected. 
It is strange that even though the column density of the ambient gas
in the hh315e region is more than the column density of the ambient 
gas surrounding HH~315B, 
we see no evidence of outflow-cloud interaction in our CO spectra. 
One possible explanation is that
the relatively high CO column density observed is all (or mostly) 
due to CO in front of HH~315E,
and that HH~315E is  interacting only with atomic gas 
{\it behind} the PV Ceph cloud.

\section{Analysis and Discussion}
\subsection{Temperature distribution}

We  use our \cooz and \coto data in concert to study the excitation
temperature ($T_{ex}$) across the mapped regions. 
We  then use the estimate of the temperature variations in our 
map to discern between
different molecular outflow entrainment models.
Since estimating the excitation temperature of optically thick ($\tau > 1$) 
gas is very unreliable (see, e.g., Figure~5 in  Hatchell et al. 1999), we
will only try to estimate the excitation temperature of optically 
thin ($\tau <1$) CO gas. And so, to obtain 
our temperature estimates
 we will use the optically thin approximation and use the following
 equation (Bachiller \& Tafalla 1999):

\begin{equation}
R_{21/10}=4  ~  e^{-11/T_{ex}},
\end{equation}
where $R_{21/10}$ is the \coto to \cooz line ratio. 

We warn that Equation 1 is only exact in the optically thin
 limit ($\tau << 1$).   
 Using Equation 1 with gas with an
opacity as low as $\tau \sim 0.1$ would lead to an underestimation
of  the real excitation temperature.
If $\tau \sim 0.1$,  the discrepancy stays within 20\% for CO 
line ratios lower than 2.5, but for line
ratios higher than 2.8, the error in $T_{ex}$ exceeds 40\%.
In any case, 
it is still true that for any given opacity, 
the higher the line ratio, the higher the excitation temperature.
So even though Equation 1 does not give  a  perfect estimate for 
$\tau \gtrsim 0.01$, we can still use it
to investigate the {\it relative} temperature distribution.

\subsubsection{The hh315b+c region}

To estimate the excitation temperature of the outflow gas in the region 
 surrounding the optical knots HH~315B and HH~315C we used
velocity-integrated maps of the \coto and \cooz lines.
For each velocity-integrated map, we produced a map of the line ratio 
($R_{12/10}$),  and then a map of 
the excitation temperature, using Equation 1.
Five grey-scale maps of $T_{ex}$  are shown in Figure~11
(the velocity ranges of integration are the same as  Figures~3{\it b} 
to 3{\it f}).
Gas with velocities close to the ambient velocity are optically thick, 
and so we do not
obtain an estimate of $T_{ex}$ for gas at  those velocities.
For each grey-scale temperature map, we superimposed the \coto 
velocity-integrated intensity contours.
We only obtain a value of $T_{ex}$ for pixels with  a signal to noise ratio of 
at least 5, in both \coto and $^{12}$CO(1--0). We masked (set $T_{ex}=0$) 
the low signal-to-noise
pixels and those (few) which gave unphysical negative values of $T_{ex}$.

In all panels of Figure~11, there is a discernible trend in which there is a 
temperature
increase in regions with high velocity outflowing CO emission.
That is, whereas $T_{ex}=10.5$~K for ambient cloud velocities (see Paper I), 
$T_{ex} > 11$~K for outflowing gas.
In addition, the average $T_{ex}$ increases with outflow velocity
(see also Table 3).
This  temperature distribution is consistent with the temperature
distribution expected for a molecular outflow  formed by bow shock
prompt-entrainment. Analytical and numerical jet-driven bow shock models
show that the temperature of the accelerated gas which forms the molecular 
outflow should be higher than the ambient gas. 
In addition, the temperature is also
expected to rise with increasing outflow velocity
(e.g., Hatchell et al.~1999;  Lee et al.~2001).

It should be noted that the warmer outflowing gas observed 
does not come directly from the
shock cooling length behind the shock front. Bow sock-driven outflow
models predict that the outflowing gas is heated as a consequence of the
acceleration (i.e., increase in kinetic energy) of the gas driven
by the momentum-conserving interaction between the bow shock and the
ambient medium (see Hatchell et al.~1999 for more on this).
Another possible source of heating near an HH object may also be the UV
radiation from the jet shock (see Wolfire \& K\"onigl 1993; Taylor \& 
Williams 1996).

\subsubsection{The hh215 region}

The \twco gas  detected (at most velocities) 
in the hh215 region  is moderately optically thick  
($\tau \sim 1$, see Paper I), 
and so estimates of the excitation temperature 
are very unreliable. 
Hence, we study the {\it relative} temperature distribution
by studying the spatial distribution of the CO line ratio ($R_{21/10}$).

The map of CO line ratio of the  blueshifted outflow gas in the hh215 region
 does not show any significant structure, and hence is not shown.
The only clear trend in this  map is that there is an increase in $R_{21/10}$ 
at the position of the source, implying there is an increase in temperature 
at the position of PV Ceph.

In Figure~12 
we plot the average line ratio along the axis of
 the redshifted molecular outflow lobe.
To do this, we determine the average value of $R_{21/10}$ 
over the width of the jet-like outflow lobe,
for each row of pixels, and then plot it as a 
function of declination offset from the source.
The line ratio shown in Figure~12 
has a maximum at the source position, then it decreases with distance
from the source up to about 36\arcsec \/ south of PV Ceph. At about 
60\arcsec \/ south of the
outflow source position $R_{21/10}$ reaches a local maximum. Further south, along
the outflow lobe axis, the CO line ratio stays approximately constant. 

Models where the molecular outflow is formed by turbulent mixing
of the ambient gas along the sides of a jet or wind predict that the gas 
temperature should
have a maximum temperature at (Cant\'o \& Raga 1991)
or very close to (Lizano \& Giovanardi 1995) the position of the source,
and decrease 
monotonically with distance 
from the source. 
Entrainment models where the gas is accelerated solely by the leading 
bow shock in a jet
predict that the gas temperature should be
 minimum at the outflow source, and increase toward the head of the 
bow shock, where the 
gas temperature peaks.  Therefore,  the gas temperature distribution 
implied by our 
measurements of $R_{21/10}$ is not entirely consistent with 
an outflow entrained by the leading jet bow shock nor by a turbulent 
mixing layer along the
sides of a jet.

One alternative explanation is that the outflow is entrained by a 
time-varying (pulsed)
jet. A time-varying jet will have internal bow shocks 
(usually called internal working surfaces)
along its axis (see, e.g., Raga et al.~1990; Stone \& Norman 1993b; 
Lee et al.~2001), and 
the gas  temperature should increase at the head of each internal shock. 
Thus, each local increase
in the CO line ratio could arise from the local increase in temperature 
expected at the head of each internal
bow shock. In this picture, the increase in line ratio 
60\arcsec \/ south of PV Ceph
could be due to the redshifted
counter-knot of HH~215(1) ---presumably also responsible for the 
``bump'' in the 
integrated intensity of the redshifted molecular outflow lobe (see Figure~12). 
We further
discuss this in \S 4.3.
 The increase in temperature at the source 
position could arise from an unresolved
internal working surface
very close to PV Ceph.

\subsection{Kinematics and momentum distribution}

Studying the velocity and momentum distribution of the molecular outflow
may help us distinguish between different entrainment models. It also helps us
better understand  how young stellar outflows interact with the ambient gas.

\subsubsection{Kinematics of the hh315b+c region}

In Figure~13 we show a \coto position-velocity ($p-v$) diagram of the
hh315b+c region. 
This $p-v$ diagram  was constructed by rotating our image of the hh315b+c 
region by 43\arcdeg \/ and
summing the \coto spectra at
 each row of pixels.

Figure~13 clearly shows that the velocity peaks at the position of HH~315B,
and decreases towards the position of the source.
This velocity distribution, where the velocity peaks at the 
shock head and decreases towards the outflow source, is consistent
with models
of bow shock-entrained molecular outflows.
For these models 
such velocity structure
is a natural consequence of the fact that the highest (radial) 
velocities are found at the head of the bow
shock, while the velocity decreases towards the wings.

Lee and coworker's
recent analytical (Lee et al.~2000)
and numerical (Lee et al.~2001) studies show   detail $p-v$ diagrams
 of wide-angle wind-driven molecular outflows. Their results indicate that 
the $p-v$ diagrams of wide-angle wind-driven molecular outflows have
noticeable differences compared with the predicted $p-v$ diagram
for  bow shock-entrained molecular outflows (see Figure~1
for a schematic illustration on this).
We do not see any indication in Figure~13 
of a structure similar to that
shown in the $p-v$ diagrams of Lee et al.~(2000; 2001) 
for wide-angle wind-driven molecular outflows. 

Figure~13 shows that the highest outflow velocities
are at the
head of the shock (i.e., HH~315B), and that the rise in velocity
associated with HH~315B is restricted to a limited region of no more than
100\arcsec \/ ($\sim 0.2$~pc). This indicates that most, if not  all, 
of the entrainment is taking
place at the head of the HH~315B shock.
 Thus, we totally discard turbulent entrainment along the sides of a jet 
(or wind) as the mechanism responsible for the outflowing gas associated
 with the HH~315B optical knot, and we suggest 
that bow shock entrainment is responsible instead.

North of HH~315B, there are two more local peaks 
in the velocity (see Figure~13).
Both of the velocity peaks come from outflowing gas associated with  
the HH~315C molecular
bow structure. The northernmost local velocity peak is coincident 
with the brightest optical emission
in HH~315C, which is also the head of the optical bow shock. 
The other local peak in velocity 
associated with HH~315C
(at $\sim 100\arcsec$ \/ north of HH~315B)
comes from outflow gas in the wings of the HH 315C CO bow
(see Figures~3{\it c} and 3{\it d}).   
As discussed above, a bow shock-entrained molecular
outflow should have the peak velocity at the head of the shock, and {\it no}
 local peak in velocity is expected at
the bow wings. The velocity structure of the outflowing gas associated with
 HH~315C does not resemble  the velocity structure of a
 wide-angle-wind-driven molecular outflow either (see Lee et al.~2000; 2001). 
Turbulent entrainment is again discarded as the head of the shock does not
show the slowest velocities in the outflow, as predicted by
 turbulent jet models. 
Although the kinematics of the molecular 
outflow associated with HH~315C are not entirely
consistent with bow shock entrainment, other 
pieces of evidence presented here suggest
that bow shock entrainment is still the best candidate 
(see \S 4.3 
for further discussion).

\subsubsection{Momentum distribution in the hh315b+c region}

In order to study the momentum distribution of the blueshifted outflowing gas 
 surrounding HH~315B and HH315C, we constructed a momentum map of the region.
The map was produced using the technique to estimate mass
described in Paper I. With this method 
we obtain a map of the outflow mass for each position pixel
and velocity channel $(x,y,v)$. We  multiply the mass at each $(x,y,v)$ 
by the (radial) outflow
velocity corresponding to the given velocity channel 
($v_{out}=v-v_{amb, north}$)
to obtain the line-of-sight (or radial) momentum at each pixel and channel.
By integrating (summing) over velocity channels
$[\Sigma m(x,y,v_i) v_{out, i} = P(x,y)]$, we 
obtain a momentum map over a given
velocity range.

We stress that we are only considering the line-of-sight (or radial) 
component of the outflow momentum. In order to obtain the true 
momentum one needs to assume an inclination angle ($i$) between the 
flow's axis and the plane of the sky. By comparing the $p-v$ diagram
of the hh315b+c region (Figure~13) with the $p-v$ diagram of modeled 
bow shock-driven outflows (e.g., Lee et al.~2000; 2001; Smith et al.~1997)
it appears that $0 < i < 30\arcdeg$ in the hh315b+c region.
In addition, the precession model of GKW estimates that HH~315 has an
inclination to the plane of the sky of about $10\arcdeg$. Thus, a value
of $i \sim 10\arcdeg$ seems adequate for HH~315. For the purpose of the 
entrainment mechanism study in this paper, it is not important to correct the
momentum by the inclination angle. However, if the true outflow momentum
(and kinetic energy) is desired we recommend that a value of  
$i \sim 10\arcdeg$ should be used.

In Figure~14
we show the momentum map of the hh315b+c region, integrated over the
velocity range 
$-15.27 < v < -0.09$~\kms. The bow-like structure of the outflowing 
gas associated with 
HH~315C is clearly seen in the momentum  map. In this structure the 
bow wings show 
more momentum than the head of the bow. In addition, the southwestern 
wing has considerably
more momentum than the northeastern bow wing. The maximum momentum in 
Figure~14
is nearly coincident  with the brightest optical emission in HH~315B, and it is
surrounded by a region of relatively high
momentum extending south and extending east of the momentum peak.
This extension traces the bow wings of the HH~315B CO bow shock, similar 
to what is seen
in the HH~315C CO bow shock, but at a smaller scale.

In a bow shock-driven outflow model
the outflowing gas velocity peaks at the head of the bow shock. 
The momentum ($p = mv$), on the other hand, is dependent on the underlying 
ambient cloud mass
distribution (Chernin \& Masson 1995). 
If a bow shock from a stellar wind mass ejection
were to interact with an ambient gas with a  perfectly uniform  
density distribution, 
the resultant
 molecular outflow  momentum  would peak at the head of the bow shock.
This is not the case for HH~315C. As discussed in \S 3.1, in this region
there is a gradient in the ambient cloud density (seen in the ambient 
CO emission, see 
Figures~3{\it j} and 3{\it k}), which increases from east to west. 
This explains why the southwest bow wing has more momentum than the 
northeast bow wings.
Near both southwest and northeast bow wings the gas is denser than the
gas at the head of the bow structure
 (we only detect \thco emission near the HH~315C CO bow wings, see Figure~5).
 Thus, the increase in momentum along the
bow wings with respect to the bow head. 
We conclude that the momentum distribution of the
 HH~315C molecular bow structure can be explained
by the bow shock entrainment of ambient
gas with a non-uniform density distribution.

The momentum distribution of the outflowing gas associated with HH~315B follows
what is expected for a bow shock-entrained outflow in a relatively flat 
ambient density distribution.
 As can be seen
in Figure~3 the outflowing gas associated with HH~315B is constrained 
to a small area 
surrounding HH~315B, and the ambient cloud density is approximately 
constant within that area.
In Figure~14,
 the momentum peaks practically at the presumed bow shock head, and the momentum
decreases  away from the head, along the wings, towards the direction 
of the outflow source.

\subsubsection{Velocity distribution in the hh215 Region}

Similar to  hh315b+c, we studied the velocity distribution of the hh215 
area by constructing
a position-velocity diagram of the \coto emission. The $p-v$, shown in 
Figure~15, was
made by summing all  spectra over the width of the hh215 area at each 
different row of pixels,
resulting in a declination-velocity diagram.

There are several interesting features in the hh215
 $p-v$ diagram.  
The redshifted gas,  shows two distinctive velocity peaks; 
one is coincident with the position of PV Ceph, and the other is 
$\sim 60\arcsec$ \/ (0.15~pc)
south of PV Ceph.
Most of the slow redshifted velocity coincident with the outflow 
source position comes from the east-west cometary-like
structure seen in Figure~6{\it c}. 
We believe that most of this structure is a result of the motion of PV
Ceph through  its parent cloud (see  Goodman \&  Arce 2002). 
So, most of the slow redshifted gas at the position of
PV Ceph is not related to the HH~315 outflow. 
Thus, we only consider the redshifted gas south of PV Ceph as
being part of the outflow.

The peak in velocity, $\sim 0.15$~pc south of PV Ceph, is coincident 
(within the 11\arcsec \/
beam of the telescope) with
the ``outflow clump'' (or  local maximum in the outflowing CO) 
south of PV Ceph,  seen
in the \coto velocity-integrated map shown in Figure~6{\it d}.
A peak in outflow velocity coincident with the position of an outflow clump
 has been
attributed, in other outflows, to be evidence for bow shock (prompt)
 entrainment from a mass ejection
episode (e.g., RNO~43, Bence et al.~1996;  HH~300, Arce \& Goodman 2001b). 

We believe the optically undetected
redshifted counter-knots of the HH~215 chain of knots (see Figure~6)
 are responsible for the entrainment of the redshifted outflow gas south of 
PV Ceph.
Each of the 3 major blueshifted  knots of 
HH~315 (knots A, B, and C)
 have a redshifted counter-knot (see Figure~2), and for each of the three 
redshifted-blueshifted
knot  pairs the distance from PV Ceph to the blueshifted knot is the same  
(within 
10\%) to the distance from
PV Ceph to the corresponding redshifted counter-knot (RBD).
One would reasonably expect the same for the HH~215 knots, 
and so we believe that the rise in velocity $\sim 0.15$~pc south of PV Ceph
is produced by the (unseen) counter-knot of HH~215(1).
It is very probable that optical observations have not detected HH knots in 
the corresponding
location south of PV Ceph because of the high extinction in this region.

As discussed in \S 3.2, we 
 detect a wide-angle blueshifted outflow emission north of PV Ceph,
 in the hh215
region. In the $p-v$ diagram (Figure~15), 
the blueshifted gas shows a 
Hubble-like  velocity distribution; the slowest  blueshifted velocity  
(detached from the
ambient cloud emission in
the $p-v$ diagram) is at $\sim 0.1$~pc north of PV Ceph and the average 
blueshifted
outflow velocity increases with distance from the source, off our map limits 
(see Figure~15). 
Given the limited coverage of the area,
we cannot conclude if such velocity distribution  is  consistent
with bow a shock-driven outflow model
or wide-angle wind-driven molecular outflow model.

\subsubsection{Momentum distribution in the hh215 region}

 As discussed in \S 3.2, 
we are unable to estimate the total mass and momentum from the 
blueshifted outflow gas just 
north of PV Ceph (in the hh215 region) because of contamination from
the emission from another cloud in the same line of sight. 
This hinders our ability to compare
the {\it total}
blueshifted momentum with that expected by different entrainment models.
 Hence, we do not 
consider it  here.

One the other hand, we can study the outflow momentum distribution of the 
redshifted lobe.
In Figure~16, we plot the average momentum along the redshifted outflow axis, 
using a momentum map integrated over the
velocity range of $3.16 < v < 6.46$~\kms. 
This plot was constructed  by calculating the average momentum at each horizontal
row of pixels along the north-south axis of the outflow, 
and then plotting the average momentum as a function
of distance from the source.
In order to avoid redshifted emission not associated with
the redshifted outflow lobe (i.e., the cometary-like east-west structure), 
we averaged the momentum
over a restricted area, which only includes the north-south redshifted jet-like 
outflow structure 
(see Figure~16).
The average momentum along the redshifted lobe axis has a maximum
at the source position and it decreases with distance from the source. 
 This momentum distribution is very
similar to that predicted  by Chernin \& Masson (1995)
for an outflow consisting of material swept-up by a jet bow shock 
traveling through an ambient cloud 
 with a density gradient proportional to $r^{-a}$,
where  to $1 \leq a \leq 2$.
In this model the outflow momentum decreases with distance
from the outflow source
 because the ambient cloud density decreases with distance from the young star.
The ambient density of the PV Ceph cloud (as implied by the integrated 
intensity \thco maps
of the cloud in Paper I)  
monotonically decreases with distance from PV Ceph in the region
where we detect the redshifted jet-like outflow feature. Therefore, 
the bow shock entrainment 
model of Chernin \& Masson (1995)  can be used to explain 
 the momentum distribution along the axis
of the redshifted outflow lobe in the hh215 region.

\subsection{Bow shocks, jets, and wide-angle winds}
The temperature distribution, the kinematics, 
the momentum distribution, and the morphology
of the outflow gas may all be used
to deduce the most likely entrainment mechanism
responsible for the molecular outflow.
In this section we  summarize our results, 
and discuss which are the most likely
entrainment mechanisms that accelerate the molecular gas of 
the outflow associated
with PV Ceph.

\subsubsection{Blueshifted CO bow shocks in HH 315C and HH~315B}

In Table~5 we list the properties of the outflow gas associated with HH~315B.
The morphology, as well as the temperature, velocity, and momentum 
distribution of
the outflow gas are all consistent with bow shock entrainment models. 
Thus, we conclude
that the outflow gas associated with HH~315B is bow shock-driven.

In Table~6 we list the characteristics of the molecular outflow gas 
associated with HH~315C. Some of the
properties listed are both consistent with bow shock and wide-angle 
wind entrainment.
As  discussed below, HH~315 is an episodic
outflow, in which HH 315C and HH~315B come from consecutive episodes. 
The mass ejection episode responsible for HH~315B entrains the ambient gas 
with a bow shock,
so the same is expected for HH~315C. 
It is highly unlikely that two consecutive mass ejection episodes interact 
with the environment through 
two different mechanisms.
Thus, we suspect that the outflow gas associated with  
HH~315B and HH~315C are both 
 produced by the bow shock entrainment of an episodic jet.

We note that there is still the possibility that the the underlying stellar 
wind that produces the outflow in this area is technically not a 
jet bow shock, 
but a very collimated
angle-dependent wind.
How a wide-angle wind interacts with the ambient gas
depends on the angular distribution of the wind force 
(e.g, Matzner \& McKee 1999, and references
therein). If the wind force is highly concentrated on the pole, 
then the wind would essentially be  jet-like.
This jet-like wind would  
interact with the ambient medium very much like a 
bona fide jet.
Thus, outflows created by 
a highly collimated angle-dependent wind and outflows created by a 
jet  show very similar  morphologies, and
velocity and momentum distributions
 ---any differences would be indistinguishable by our observations.
Thus, a very collimated angle-dependent wind,  
and a bona fide jet  will produce practically the same entrainment  mechanism. 

\subsubsection{The redshifted CO jet south of PV Ceph}

The properties of the redshifted molecular outflow lobe 
(south of PV Ceph in the hh215 region) are
listed in Table~7.
The redshifted outflow gas in this region shows a clear jet-like 
(very collimated) structure
that extends south of PV Ceph (see Figure~17).
Turbulent entrainment has been proposed as an attractive model to 
explain other molecular
 outflows with similar highly collimated morphologies 
(e.g., HH211, Gueth \& Guilloteau; and NGC 2024,
Richer et al.~1992). But, as can be seen in Table~7, 
the morphology is the only characteristic
of the redshifted outflow lobe that is consistent with 
turbulent jet entrainment models.

Bow shock entrainment by a variable jet with internal working surfaces
seems to better explain our results.
A major internal working surface at $\sim 60\arcsec$ \/ (0.15~pc) 
south of PV Ceph may naturally explain all of the
following characteristics which are seen there: 
1) the local increase in outflow gas column density 
(i.e., outflow clump or hot spot) seen in Figure~6{\it d}; 
2) the slight rise in the CO line ratio (Figure~12); 
and  3) the peak in outflow velocity
(Figure~15). 
We thus conclude that the redshifted
outflow gas south of PV Ceph has most likely been entrained 
by a variable jet with internal bow shocks.

The chain of optical HH knots that form HH~215 is also highly 
suggestive of a variable jet morphology.
The average axis of  HH~215 is coincident with the general 
north-south axis of the redshifted molecular
outflow lobe south of PV Ceph (see Figure~17). 
The HH~215 knots show a wiggling pattern
somewhat similar to the redshifted outflow 
lobe axis (see Figure~17).
In addition, HH~215(1)  ---the brightest knot in HH~215---
is $\sim 55\arcsec$ \/ north of PV Ceph. 
Thus, we strongly believe that the redshifted molecular outflow 
is entrained  by the counter jet of HH~215, and that the features
observed $\sim 60\arcsec$
\/ south of PV Ceph are produced by the counter-knot of HH~215(1).

It is interesting to note that although the outflow gas 
in the hh315b+c region (Figure~3)
and the redshifted outflow in the hh215 region, south of PV Ceph, 
(Figure~17) are both 
presumably entrained by the same mechanism, 
they have very different morphologies.
 The source of this seeming
inconsistency is apparent from the morphology of the
optical HH~315 flow.
Similar to other HH flows [e.g., HH~34 
(Reipurth et al.~1986; Devine et al.~1997);
and HH~111 (Reipurth et al.~1992; Reipurth et al.~1997; RBD)]
the HH objects which make up the HH ~315, 
increase in size, the further away they are from the
outflow  source. 
We should expect the redshifted counter-knots of HH~215, 
south of PV Ceph, to be
as compact and small as HH~215. Therefore, the
redshifted gas  south of PV Ceph has a jet-like 
appearance because it is most probably 
entrained by a continuous (unresolved) 
chain of  small bow shocks, which have a jet-like appearance.
 On the other hand, the transverse
size of
 the bow shock responsible for the
blueshifted outflow associated with HH~315C is  
larger than the shocks associated
with HH~215, 
and so the CO outflow bow shock-like structure 
produced is  very well resolved by our observations. 

\subsubsection{Evidence for a  blueshifted wide-angle wind north of PV Ceph}

The blueshifted \twco and \thco gas just north of PV Ceph, in the hh215 region,
shows a wide-angle structure. The \thco integrated emission
has a V-like structure which encloses the 
blueshifted emission seen in $^{12}$CO (see Figure~8). In addition, 
the \thco V-structure coincides with the 
walls of the optical conical reflection nebula. Thus, we are confident 
that  the observed blueshifted
\thco  structure traces the limb-brightened walls of a 
wind-blown cavity.
This structure is similar to that observed in the inner regions of the B5-IRS1 
outflow (Langer, Velusamy, \& Xie 1996; Velusamy \& Langer 1998; 
Yu et al.~1999, hereafter YBB).

Approximately coincident with the cavity axis, lie the HH~215 chain of knots.
These optical knots trace  gas that has recently been
excited by a very collimated wind ---which seems to be very different
from the wide-angle wind responsible for the V-shape cavity.
One explanation which could explain the observations is that
there are two different wind components in the northern 
(blueshifted)
lobe close to
PV Ceph: 1) a collimated (jet-like) wind responsible for the HH~215 knots; 
and 2) a wide-angle wind responsible for the poorly collimated 
blueshifted $^{12}$CO immediately north of PV Ceph,  and the \thco V-shaped structure.
A similar two-component wind is observed in the 
inner region of the B5-IRS1  outflow (see YBB).

There is evidence that a ``dual wind component'' is also present 
in the southern (redshifted) lobe
of the outflow associated with PV Ceph.
 As discussed above, the collimated outflow lobe south of PV Ceph is most
probably produced by a variable jet (the counter jet of HH~215). 
We do not
detect a wide-angle wind in the redshifted lobe south of PV Ceph, 
but optical observations of
the region close to PV Ceph (Gledhill et al.~1987; Levreault \& Opal 1987; 
Neckel et al.~1987;
RBD) have detected a fan-shaped reflection nebula south of PV Ceph, 
similar to the reflection
nebula north of PV Ceph (see Figure~8). 
It is very probable that the redshifted molecular gas emission of the
wide-angle wind south of PV Ceph is ``hidden'' under the 
optically thick ambient cloud emission
of the region.

If the above picture is correct, a model which allows for the coexistence
of both a wide-angle and a very collimated wind is needed to explain 
the outflow from PV Ceph, similar
to the B5-IRS1 outflow (YBB). So, 
 we believe, as YBB do for B5-IRS1, that
the sum of all the observations of 
the PV Ceph outflow near the source  can  be best described 
with a two-component wind model (like that of Hirose et al.~1997)
or a single-wind model in which
the wind splits into a very collimated (axial)  
component and a wide-angle component
(like the X-wind model of Shu et al.~1995, 
and references therein; see also Shang, Shu, \& Glassgold 1998). 

Alternatively, as proposed to us by the anonymous referee, the wide-angle 
cavity could be a result of the episodic and precessing nature of the HH~315
flow. Sideways splashing from each HH flow episode could, in principle,
slowly burrow through the ambient gas, creating a wide-angle cavity without
the need of invoking another  (wide-angle) wind component.
Although it is possible
that a wide-angle cavity could be formed in such a way,
we find that this picture does not entirely fit the observations of PV Ceph.
For example,
sideways splashing from the HH~315C,B,A episodes
would have helped formed a cavity with an axis 
tilted towards the west, unlike the observed 
nice V-shape cavity with an almost perfect north-south axis. 
Also, the V-shape cavity is unlikely to have been produced 
solely by the sideways splashing of the
HH~215 episode since the cavity extends more towards the
north than the HH~215 chain of optical knots. 
Thus, we prefer a wide-angle wind
component in order to explain the existing observational data. 
Further kinematic
studies of the molecular gas near PV Ceph should help clarify the
nature of the wide-angle cavity.

\subsection{Episodicity and axis wandering of the HH~315 flow}

\subsubsection{Episodicity}

Previous optical studies  have pointed out the possible   episodic
 nature of the HH~315 giant HH flow (GKW, RBD). The optical evidence
for the episodicity of HH~315 comes from the fact that each of the three
 major HH knots in each lobe is about 0.35~pc from each other, with no
HH-like emission between them.
An HH knot is produced by the shock arising from the interaction
of a high-velocity  flow of gas ejected by a young star and the ambient medium.
 In HH~315, the HH knot pairs
C-F, B-E, and A-D (see Figure~2) 
are thought to arise from three different mass ejection episodes.

Our millimeter line data shows further evidence for the episodic  
nature of the 
HH~315 flow. The outflow gas surrounding the HH~315B and HH 315C knots
has a spatially discrete structure (Figure~3)  which  shows a 
shell-like or bow-like structure at the
position of each of the two knots. At the head of each of these 
shell-like structures there is an
increase in outflow velocity (Figure~13). 
A velocity increase at the position of the optical HH knots A, B, C, 
and D is also 
observed in the large-scale \coto $p-v$ diagram of the HH~315 outflow 
(Figure~13 in Paper I).
Such  morphology and velocity distribution in the molecular
outflow gas is not expected
if the underlying stellar wind responsible for the creation of the
 molecular outflow were made of
a continuous constant flow of ejected mass.
The multiple CO shell structure, and the peak molecular outflow velocity 
at the head of the shock (the position of the HH knot)
 is better understood if the molecular outflow from PV Ceph
 is formed by
a wind with sporadic episodes of copious mass loss 
(see also Arce \& Goodman 2001a).

\subsubsection{Wandering ejection axis}

The fact that the ejection axis of HH 315 changes over time has been well 
established by the optical images
of the flow, where it is clearly seen that the 
HH knots trace an S-shaped path.
Tracing a line from PV Ceph to the location of each knot, it is seen
that each major knot has a different position angle on the sky. 
Each mass ejection should travel in a
straight line after being ejected by the young star, 
unless it collides with a dense clump which could
change its trajectory. 
Our large scale molecular gas maps (Paper I)
do not show any evidence for
dense clumps which could have perturbed the ballistic 
trajectory of any ejection.
Thus, the HH knots in HH~315 are at different angles with 
respect to PV Ceph, because
PV Ceph's angle of
ejection is changing over time (i.e., wandering or precessing).
GKW have successfully reproduced the morphology of 
the HH~315 flow with a simple precession model,
assuming a jet velocity of 200~\kms, a precession 
cone with full opening angle of $\sim 45\arcdeg$,
an inclination to the plane of the sky of $\sim 10\arcdeg$, 
and a precession period of about
8300 yr. Goodman \& Arce (2002) show that a westward
 motion of PV Ceph, in addition to a time-varying
ejection angle, explain the position of the HH knot pairs 
(A-D, B-E, and C-F) with respect to the source better than models
without transverse source motion.

In addition to the ``precession'' (or change in ejection angle)
 traced by the optical HH knots, our millimeter CO data for the
redshifted outflow gas in the hh215 region show signs of  
a small scale time-varying ejection angle. 
 In Figure~17 we show the 
integrated intensity contours of the redshifted 
gas in the hh215 region, integrated
over the velocity range where we detect outflow emission. 
We made east-west intensity cuts for all
pixel rows, for the extent of the north-south (jet-like) 
outflow structure. 
Most of the cuts show Gaussian-like intensity profiles, 
so we fit a Gaussian to each cut
and obtained  the position of the Gaussian centroid from each fit.
In Figure~17, the thick black solid
line plotted over the intensity contours  indicates the position 
of the integrated emission centroid (obtained from the Gaussian fit)
 along
the length of the redshifted CO jet-like structure. 
Assuming the emission centroid indicates the position 
of the outflow axis, we can state
that the axis varies in direction over time (it wiggles).
Similar wiggling morphology 
has been detected in optical HH jets 
(e.g., Heathcote et al.~1996) and other
molecular outflows (Davis et al.~1997).

In Figure~18{\it a} we plot the presumed trajectory of the jet axis, 
traced by the \coto
velocity-integrated intensity centroid.
It can be seen in Figure~18{\it a} that the points 
trace a sine-like path with a slope.
We fit the centroid path with a straight line, 
and we then subtract the fit to the points, and show the result
in Figure~18{\it b}. A sine wave was subsequently 
fitted to the slope-corrected points 
(see Figure~18{\it b}).
Notice that the points at the peaks and valleys (of the sinusoidal trace) 
increasingly deviate from
the sine fit, the furthest away from the source. 
This behavior, where the axis traces 
 a cone (projected on the plane of the sky) rather than a 
cylinder,
is suggestive of a ``precession cone''
expected for a flow with a quasi-periodic time-varying ejection axis 
(due to pure periodic precession or
to a quasi-periodic random wandering of the ejection axis).

Assuming the sinusoidal appearance of the redshifted lobe  
comes from a purely periodic precession
in the ejection axis, we may use the ``wavelength'' 
($\lambda \sim 28,820$~AU)
of the sine fit made to the
points in Figure~18{\it b} to obtain a precession period. 
If we assume  a jet velocity of 200~\kmss  (the same as GKW and RBD), 
we  then obtain a
precession period ($T=\lambda/v_{jet}$) of $\sim 680$~yr. 

Tidal interactions between binary components 
and a circumstellar disk is one possible mechanism
which could induce precession on a young star.  
Terquem et al.~(1999) give an approximate expression for the
precession period of a disk around a young star in a binary system, where the
disk surrounds only the primary star. The expression is given in terms
of the primary mass ($M_p$), the mass ratio of the two stars,
the  disk radius ($R$), and the
binary separation ($D$) .
Observations show that 
if PV Ceph has a binary companion it must be less than 50~AU apart
(Leinert et al.~1997). 
We can use Equation 1 in Terquem et al.~(1999) 
and solve for the binary separation,
to see if a precession period of $\sim 680$~yr 
is possible for a binary system with a
separation of less than 50~AU between its members. 
If we assume PV Ceph is the primary star with a mass
of 4~M$_{\sun}$ (Fuente et al.~1998a), a primary 
to secondary mass ratio of about 0.25,
and a 15~AU disk radius, then a binary separation of 
$\sim 21$~AU would be needed to drive
a precession with a period of 680~yr. Even if 
we were to change our assumptions to
a primary-to secondary mass ration of 1 and a disk radius of 25~AU, 
the binary separation would be
43~AU. Thus, it is possible that the wiggling of the jet is due
 to precession of the outflow source
induced by tidal interactions between PV Ceph, 
a yet undetected binary companion,
and PV Ceph's circumstellar disk.

The  precession period of $\sim 680$~yr, from our observations of the
wiggling redshifted CO outflow lobe, is about a factor of 12
less than the precession period obtained by GKW (from modeling the trajectory
traced by the optical HH knots in HH~315). 
This apparent difference in precession period seems to indicate
that: 1) the  precession period is changing over time;
or 2) there are two different 
mechanisms which are responsible for the different apparent
precession-like motions.  

Theoretically,
the ejection of a third companion in a hierarchical 
triple system could lead to the formation
of a tighter binary (see Reipurth 2000 for more details).
But, in the case of PV Ceph the tightening 
of the binary system would  had  to occur
in a very short time-scale of  $2000$~yr.\footnote{\footnotesize
This is the approximate
time-scale between the eruptions responsible for the HH~215 chain of knots
and the HH~315A-D knot pair 
(see Figure~2), assuming $v_{jet} \sim 200$~\kms.} 
In addition, it is extremely coincidental
that we would be observing PV Ceph right 
at the moment after the tightening of the binary system.
Thus, it is unlikely that this scenario applies to PV Ceph.
It seems more likely then, that the large-scale axis wandering
is due to some other (unknown) mechanism.
Some other possible mechanisms which could 
produce precession-like motions of the outflow ejection axis are:
changes in the outflow source's magnetic field orientation; 
or a precession-like motion induced by tidal interactions 
of {\it multiple} stellar companions and their circumstellar disks.

In summary, we may explain the short-time scale
axis wandering observed in the HH~315 outflow
by circumstellar disk precession. However, it is highly unlikely that the
same mechanism is responsible for the large-scale axis wandering.

\section{Summary}

We observe, at high velocity and spatial resolution, the molecular gas 
surrounding 
several knots of the giant  Herbig-Haro flow HH~315, 
from the young star PV Ceph.
 The observations were aimed at
studying the interaction between the HH flow and the ambient gas. 
The data obtained include
simultaneous observations, at the IRAM 30 m telescope, of the
$^{12}$CO(1--0), $^{12}$CO(2--1), and $^{13}$CO(1--0), 
molecular lines.
The three regions observed include:  1)
an area surrounding two blueshifted knots (HH~315C
and HH~315B) about 0.9 to 1.2~pc northwest of PV Ceph; 
2) an area which includes the gas surrounding the
outflow source, the HH~215 blueshifted optical 
knots  $\sim$ 0.05 to 0.15~pc 
north of PV Ceph, and the
collimated redshifted molecular gas south of PV Ceph;
and 3) an area surrounding the redshifted optical
knot HH~315E, about 0.9~pc southeast from the outflow source.
The main points derived from our study can be summarized as follows:

1) We find that  the blueshifted outflow gas in 
the region surrounding  HH~315B and HH~315C
has clearly been
accelerated by bow shock entrainment of an episodic jet. The molecular 
outflow gas shows a spectacular bow-shock-shaped morphology, which  has a 
width of about 0.4~pc
at the slowest outflow velocities. 
The head of the CO outflow  bow structure coincides with the position of  
the bow-shaped optical knot HH~315C.
There is also blueshifted molecular outflow gas coincident with the optical 
knot HH~315B which
exhibits a structure suggestive of a bow shock morphology.  
A bow-like structure coincident with each knot, and the fact that
both the excitation temperature  and the 
velocity of the outflow gas show peaks at the position of the
head of  each bow-like  structure, are all consistent with 
an outflow formed by two different bow shocks.
Each of these two bow shocks were  formed by a different
mass ejection episode, where HH~315C is the ``leading jet head'' 
of the HH~315 giant flow, and 
HH~315B is a shock formed by a  subsequent mass ejection episode.

2) Near PV Ceph (within 0.3~pc), the observational
data is highly suggestive of the coexistence of a 
wide-angle wind and a 
collimated (jet-like) wind.
The blueshifted \thco integrated emission shows a 
 V-shaped morphology with an opening angle of $\sim 90\arcdeg$, 
which is coincident with an optical reflection nebula.
The blueshifted \twco has a fan-like morphology which fills the 
cavity delineated
by the  \thco V-shape structure.
 We suggest that the \thco traces the limb-brightened walls of
a wide-angle wind-blown cavity. 
Along the axis of the wide-angle blueshifted \twco outflow  
lie the previously detected optical knots HH~215. These small 
knots delineate the
collimated component of the blueshifted wind north of PV Ceph.

3) We find that the redshifted molecular outflow lobe south of 
PV Ceph is most 
likely entrained by a variable jet with several internal working surfaces 
(bow shocks).
 The redshifted outflow has a collimated wiggling jet-like appearance, 
with an ``average''
north-south axis, which extends to about 0.3~pc south of PV Ceph.
The momentum distribution 
is  consistent with jet bow shock entrainment in  an ambient medium with 
density decreasing with distance from the source. Also the velocity and 
temperature distribution
of the molecular outflow gas are consistent with  it being entrained by a 
jet with several
internal bow shocks.
We show that the same (bipolar) 
mass ejection episode responsible for the 
blueshifted HH~215 optical 
 knots (north of PV Ceph) is also responsible for the entrainment of the 
redshifted outflow lobe. 
There has been no optical detection of the redshifted counter-jet of HH~215
because of the heavy extinction in the region south of PV Ceph.

4) We find that the wiggling observed in the redshifted 
CO outflow lobe near PV Ceph
 is most probably due to a time-varying ejection axis. 
Assuming the wiggling is due to precession
of the outflow source, and a jet velocity of 200~\kms, 
then the precession period is 680~yr.  
This is about a factor of 10 less 
than the precession period deduced from the large-scale
optical HH flow. 
We may explain the short time-scale
axis wandering observed in the redshifted molecular outflow lobe
by precession induced by tidal interactions between
(undetected) binary companions and a circumstellar disk.
 However, it is highly unlikely that the
same mechanism is responsible for the large-scale axis wandering.

5) We do not detect any 
outflow emission in the region surrounding the HH~315E optical knot. 
The CO spectra in this
region has a FWHM width of only $\sim 1$~\kms, and shows no 
evidence of outflow-ambient gas
interaction. It is puzzling to find such a narrow CO width 
in a region presumably affected by a stellar outflow. It is probable that all
 of the CO observed at the direction of HH~315E is in front
of HH~315E (on the same line-of-sight), and that HH~315E is 
interacting mainly with atomic gas
behind the molecular cloud.

\acknowledgements
We would like thank John Bally, and 
Charlie Lada for their helpful comments on this work.
And we are grateful to the National Science Foundation 
for supporting
this effort through grants AST 94-57456 and AST 97-21455.

\clearpage

\figcaption[figure1.eps]{
Molecular outflow properties predicted by different entrainment models.
The rows, starting on the top, show the turbulent jet, jet bow shock, and
wide-angle wind properties. The columns, starting from the left, show a 
schematic picture of the
stellar wind, and the  model-predicted 
molecular outflow morphology, velocity profile,
temperature profile, and momentum profile. An underlying density distribution
of $r^{-1}$ to $r^{-2}$ is assumed for the momentum profiles shown. 
There are no explicit estimates of the outflow gas temperture, as a function
of distance from the source, for the wide-angle wind-driven models. References
for the turbulent jet model 
 properties shown here are: Bence et al.~(1996); Cant\'o \& Raga (1991); 
Chernin \& Masson (1995). References for the jet bow shock model 
 properties shown here
are: Lee et al.~(2001); Hatchell et al.~(1999); Chernin \& Masson (1995); 
Cliffe et al.~(1996). References for the wide-angle model properties shown
here are: Lee et al.~(2001); Li \& Shu (1996).}

\figcaption[figure2.eps]{
Wide-field H$\alpha$+[S~II] (optical) CCD image of 
the HH~315 giant HH flow, from RBD. 
The dashed boxes denote the areas mapped with on-the-fly mapping 
at the IRAM~30~m telescope.
The name given to each region is shown at the bottom of each region.
We also show selected contours of the large-scale  blueshifted (grey) and 
redshifted (black) \coto outflow gas (based on Figure~1 of Paper I). 
In certain places the contours are cut so that features in the optical image
may be seen better.
The position of the HH knots, and the
position of PV Ceph (the outflow source) are also shown.
\label{otfmaps.iram}}

\figcaption[figure3.eps]{Velocity-range-integrated
 intensity maps of the \coto emission surrounding the 
blueshifted optical knots HH~315C and HH~315B 
(i.e., the hh315b+c region, see Figure~2).
The velocity range of integration is shown 
in the upper-left corner of each panel. 
The  starting contour and the contour steps  
are given in brackets at the lower-right corner of each panel 
in units of K~\kms.
The grey silhouettes represent the S[II] knots in the region, from GKW. 
The optical knots are identified in panel  [{\it a}].
Panel [{\it a}] also shows the IRAM~30~m beam at the \coto frequency 
(11\arcsec), and the linear scale
assuming a distance to PV Ceph of 500~pc. In panel {\it g} we show 
the position from where the spectra, shown
in Figure~4, are taken. Each letter in panel [{\it g}] represents the position 
of the spectrum shown in the panel, in Figure~4, with the same letter.
\label{co21vel1.iram}}

\figcaption[figure4.eps]{Sample \coto (black) and \thcooz (grey) spectra 
of the hh315b+c region. 
Panels [{\it a}] through [{\it g}] show spectra from the position shown in
 Figure~3{\it g}.
Each spectrum ({\it a} through {\it g})
was taken from a single 10.5\arcsec \/ by 10.5\arcsec \/ pixel, 
of a molecular line
map convolved with a 21\arcsec \/ beam. Panel [{\it h}] shows the 
average spectra
over the whole hh315b+c region. The dotted vertical lines indicate the 
approximate range of the ``ambient''
cloud velocities ($1.0 < v < 2.5$~\kms) in the hh315b+c region.
\label{spechh315b+c.iram}}

\figcaption[figure5.eps]{Velocity-range-integrated
 intensity maps of the \thcooz emission surrounding the blueshifted optical knots HH~315C and HH~315B.
The velocity range of integration is shown at the top of each panel. 
The grey silhouettes represent the S[II] knots in the region, see Figure~3.
All panels have the same staring contour  and  contour step value of  0.18~K~\kms.
The linear scale is shown in panel [{\it a}] and the IRAM~30~m beam at the \thcooz frequency (22\arcsec) 
is shown in panel [{\it b}].
\label{13cohh315b+c.iram}}

\figcaption[figure6.eps]{Velocity-range-integrated
 intensity maps of the \coto emission surrounding PV Ceph, HH~215,
and the southern redshifted outflow lobe (i.e., the hh215 region, 
see Figure~2).
The velocity range of integration is shown at the top of each panel.
The starting contour and the  contour steps
are given in brackets at the lower-right corner of each panel in units of K~\kms.
The star symbol denotes the position of the outflow source, PV Ceph, and the
crosses denote the position of the different HH~215 knots 1 through 5 (from GKW),
identified in panel [{\it a}]. 
The IRAM~30~m beam at the \coto frequency
is also shown on panel [{\it a}].
 The letter inside the panels identify the position from
which we obtain the spectra shown in Figure~9. 
Each letter  represents the position 
of the spectrum shown in the panel (in Figure~9)
 with the same letter.
\label{co21hh215.iram}}

\figcaption[figure7.eps]{Velocity-range-integrated
 intensity maps of the \thcooz emission in the hh215 region.
The velocity range of integration is shown at the top of each panel.
The starting contour and the contour steps
are given in brackets at the lower-right corner of each panel in units of K~\kms.
The star symbol denotes the position of the outflow source, PV Ceph, and the
crosses denote the position of the different HH~215 knots 1 through 5 (from GKW),
identified in panel [{\it c}]. 
The  IRAM~30~m beam at the \thcooz frequency
is shown on panel [{\it b}]. The letters inside the panels identify the position from
which we obtain the spectra shown in Figure~9. 
Each letter  represents the position 
of the spectrum shown in the panel (in Figure~9) 
with the same letter.
\label{13cohh215.iram}}

\figcaption[figure8.eps]{({\it Top}) 
Velocity-range-integrated intensity contour map of blueshifted \coto
emission (within 0.3~pc north of PV Ceph in hh215 region)  superimposed on
grey-scale map of blueshifted \thcooz emission of the same area. 
The velocity range of integration and the contours of the
\coto map are the same as Figure~6{\it b}, and the grey-scale map comes from 
Figure~7{\it a}. 
The star and cross symbols show the same as previous figures.
The dashed dark line represents the eastern edge of 
the optical nebula north of PV Ceph.
({\it Bottom}) $I_c$ image of the PV Ceph biconical  
nebulosity (from Levreault
\& Opal 1987). The field shown is 2.58\arcmin \/ high.
\label{widehh215.iram}}

\figcaption[figure9.eps]{Sample
 \coto (black) and \thcooz (grey) spectra of the hh215 region. 
Panels {\it a} through {\it g} show spectra from the positions shown in Figures~6 and 7. 
The spectra shown in panel [{\it d}]
comes from the position of the outflow source, PV Ceph.
Each spectrum ([{\it a}] through [{\it g}]) 
was taken from a single 7\arcsec \/ by 7\arcsec \/ pixel, 
of a molecular line
map convolved with a 14\arcsec \/ beam. 
In panel [{\it h}] we show the average spectra
over the whole hh215 region. 
The dotted vertical line indicates the position of $v=2.5$~\kms.
\label{spechh215.iram}}

\figcaption[figure10.eps]{Average
\coto spectrum of the mapped area surrounding the
redshifted optical knot HH~315E (i.e., the hh315e region, see Figure~2).
A dashed line indicates the position of $v = 2.5$~\kms.
The main component, with a peak in $T_{mb}$ at $v = 2.5$~\kms, is 
due to ambient
gas from the cloud associated with PV Ceph. The bump in the spectrum 
at blueshifted
velocities is due to emission from another cloud on the same line of sight.
\label{spechh315e.iram}}

\figcaption[figure11.eps]{Grey-scale excitation 
temperature maps of the CO outflow gas in the
hh315b+c region. Superimposed on the grey-scale maps, we show
in  contours, the \coto velocity integrated intensity. The velocity 
range of integration
is shown at the top of each panel. The first contour  and  
contour steps for each panel are the
same as for the corresponding velocity range in Figure~3. 
The average excitation temperature 
($\bar{T}_{ex}$) for each  velocity range (from Table 3) 
is shown on the bottom-right corner
of each panel. Each pixel is 12\arcsec \/ by 12\arcsec.
\label{temphh315b+c.iram}}

\figcaption[figure12.eps]{({\it Left}) 
 \coto integrated intensity contours of the redshifted emission near PV Ceph
(hh215 region). The velocity range of integration is $3.16<v<6.46$~\kms. 
The  starting
 contour and contour step
are 2.64 and 0.88~K~\kms, respectively. The star symbol indicates the 
position of PV Ceph, 
and the crosses indicate
the position of the HH~215 knots. ({\it Right}) 
Average \coto to \cooz line ratio as a function of distance from the
source. The line ratio is averaged over the width of the north-south 
jet-like structure, 
indicated by the dark vertical dashed lines
on the left panel. The error bars indicate the 1-$\sigma$ error.
\label{temphh215.iram}}

\figcaption[figure13.eps]{
\coto position-velocity diagram of the hh315b+c region. The $p-v$ diagram 
was constructed by rotating our \coto map of the hh315b+c region 
(with 12\arcsec \/ by 12\arcsec \/ pixels, and
0.22~\kms-wide velocity channels)
by 43\arcdeg \/ and summing the \coto spectra at each row of pixels. The 
horizontal lines
 denote the position 
of the brightest optical emission in the HH~315B and HH~315C  knots.
Contours are 3 to 51 in steps of 2~K, and 56 to 91~K, in steps of 5~K.
\label{pvhh315b+c.iram}}

\figcaption[figure14.eps]{
Map of the radial component of the  molecular outflow gas momentum 
in the hh315b+c region, over
the velocity range $-15.27<v<-0.09$~\kms.
The  first contour  and contour step are 
3 and 1.5 $\times 10^{-3}$~M$_{\sun}$~\kms, respectively.
The crosses denote the position of the brightest optical emission 
of the HH~315B and HH~315C knots.
\label{momhh315b+c.iram}}

\figcaption[figure15.eps]{
\coto position-velocity diagram of the hh215 region. 
The $p-v$ diagram was constructed by summing all  spectra 
over the width of the hh215 area at each different row of pixels,
resulting in a declination-velocity diagram. 
The map used has 12\arcsec \/ by 12\arcsec \/ pixels and a 
velocity resolution of 0.11~\kms.
The horizontal line indicates the position of the outflow source (PV Ceph).
Contours are 7 to 35~K in steps of 2~K, and 40 to 80~K in steps of 5~K. 
\label{.iram}}

\figcaption[figure16.eps]{
({\it Left}) Same as  left panel in Figure~12.
({\it Right}) Average momentum in the redshifted outflow lobe  as a function of
distance from PV Ceph.
The momentum was averaged over the width of the redshifted CO jet-like 
structure, indicated by the
black vertical dashed lines on the left panel.
\label{momhh215.iram}}

\figcaption[figure17.eps]{\coto integrated intensity contours of the wiggling 
molecular 
redshifted outflow lobe in the hh215 region, 
near PV Ceph. The contours are the same as the left panel of 
Figure~12. 
The emission centroid (obtained from Gaussian fits to the intensity
profile)
is indicated by the thick black line.
The star symbol represents the position of PV Ceph. The crosses represent the 
position of the HH~215(1) through HH~215(5)
 knots, and the number besides each cross indicates the HH~215 knot number.
\label{wiggling.iram}}

\figcaption[figure18.eps]{
({\it a})  Plot of the redshifted outflow lobe emission centroid position.
Distances are given in terms of pixel numbers and arcseconds from the map
 (Figure~17) edges.
The straight line fit to the points is plotted, and the fit parameters 
are shown. 
 ({\it b}) Plot of the redshifted outflow lobe emission centroid position, 
corrected for the slope indicated by the line fit in [{\it a}]. The
sinusoidal fit to the points, and the resultant fit parameters are shown.  
Axes are in arcseconds offsets and distance in AU from the
source. In both panels, the  errors shown are 3-$\sigma$ errors.
\label{wig_fit.iram}}

\clearpage

\begin{deluxetable}{lccc}
\tablewidth{0pt}
\tablecolumns{4}
\tablecaption{Regions Mapped with the IRAM 30~m telescope
\label{otfiram}}
\tablehead{
\colhead {Region} &
\multicolumn{2}{c}{Center Position} &
\colhead {Region}\\ 
\cline{2-3}   
\colhead{Name} &  
\colhead{$\alpha_{1950}$} &
\colhead{$\delta_{1950}$} &
\colhead{Size} 
}
\startdata

hh315b+c & $20^h44^m49^s.4$ & $67\arcdeg52\arcmin49\arcsec$ & $5.2\arcmin 
\times  4.2\arcmin$\\
hh215 & $20^h45^m26^s.4$ & $67\arcdeg46\arcmin13\arcsec$ & $2.2\arcmin  
\times  4.2\arcmin$\\
hh315e & $20^h45^m59^s.6$ & $67\arcdeg40\arcmin31\arcsec$ & $2.3\arcmin  
\times  2.3\arcmin$
\enddata  
\end{deluxetable}

\clearpage

\begin{deluxetable}{lccccc}
\tabletypesize{\footnotesize} 
\tablewidth{0pt}
\tablecolumns{6}
\tablecaption{RMS Noise and Velocity Resolution of Spectral Line Maps 
\label{rmsmapsiram}}
\tablehead{
\colhead{Region} &
\colhead{Molecular} &
\colhead{Beam\tablenotemark{a}} &
\colhead{$\delta v$\tablenotemark{b}} &
\colhead{RMS\tablenotemark{c}} &
\colhead{Where}\\
\colhead{Name} &
\colhead{Line} &
\colhead{[\arcsec]} &
\colhead{[\kms]} &
\colhead{[K]} &
\colhead{Used}
}
\startdata
hh315b+c & \coto & 21 & 0.22 & 0.62 & Fig.~3 \\
                & \thcooz & 21 & 0.22 & 0.20 & Fig.~5\\
                & \coto & 24 & 0.22 &  0.50 & Figs.~11, 13\\
                & \cooz & 24 & 0.22 & 0.28 & Figs.~11, 14, 
Mass estimates\tablenotemark{d}\\
                & \thcooz & 24 & 0.22 & 0.16  & Mass estimates\tablenotemark{d}\\
hh215      & \coto & 14 & 0.11 & 0.78 & Fig.~6 \\
                & \thcooz & 14 & 0.11 & 0.28 & Fig.~7\\
                & \coto & 24 & 0.11 &  0.49 & Fig.~12\\
                & \cooz & 24 & 0.11 & 0.31 & Figs.~12, 16,
Mass estimates\tablenotemark{e}\\
                & \thcooz & 24 & 0.11 & 0.18  & Mass estimates\tablenotemark{e}\\
hh315e    & \coto & 21 & 0.11 & 0.69 & Fig.~10
\enddata

\tablenotetext{a}{Size of OTF map convolution beam. The size
of each pixel in each map is always half the size of the convolution beam.}
\tablenotetext{b}{Velocity resolution (width of velocity channel).}
\tablenotetext{c}{Maximum RMS (in $T_{mb}$ units) of  all
pixels {\it not} at the edge of the map.}   
\tablenotetext{d}{Outflow mass estimates of the hh315b+c region are shown in 
  Table~3}
\tablenotetext{e}{Outflow mass estimates of hh215 region are shown in 
  Table~4}
\end{deluxetable}

\clearpage

\begin{deluxetable}{ccccc}
\tabletypesize{\footnotesize} 
\tablewidth{0pt}
\tablecolumns{5}
\tablecaption{Mass at different velocity ranges in hh315b+c
\label{masshh315b+c}}
\tablehead{
\colhead{Velocity Range\tablenotemark{a}} &
\colhead{$\bar{T}_{ex}$\tablenotemark{b}} &
\colhead{Mass} &
\colhead{Momentum\tablenotemark{c}} &
\colhead{Kinetic Energy\tablenotemark{c}} \\
\colhead{[\kms]} &  
\colhead{[K]} &
\colhead{[M$_{\sun}$]} &
\colhead{[M$_{\sun}$~\kms]}  &
\colhead{[$10^{43}$ erg]}
}
\startdata

$-15.27 < v < -10.43$ & 21.7 &  0.01 & 0.14 & 1.94\\
$-10.43 < v < -6.03$  & 18.6 &  0.03 & 0.25 & 2.27\\
$-6.03 < v < -3.39$   & 17.8 &  0.05 & 0.30 & 1.76\\
$-3.39 < v < -2.29$   & 15.4 &  0.07 & 0.30 & 1.26\\
$-2.29 < v < -1.19$   & 13.0 &  0.17 & 0.53 & 1.64\\
$-1.19 < v < -0.09$   & 11.0 &  0.76 & 1.52 & 3.10\\
$-0.09 < v < 1.01$    & 10.5 &  2.88 & 2.65 & 2.69\\
$1.01 < v < 1.89$\tablenotemark{d} & 10.5 &  3.92 & \nodata & \nodata\\
$1.89 < v < 2.55$\tablenotemark{d} & 10.5 & 1.31 & \nodata & \nodata\\
$2.55 < v < 3.65$     & 10.5 &  0.58 & 0.77 & 1.08 
\enddata  
\tablenotetext{a}{Velocity ranges are the same ranges as the ones in Figure~3.}
\tablenotetext{b}{Average excitation temperature at the given velocity range, using 
Equation 1, except for last four rows where $\bar{T}_{ex}=10.5$ (from Paper I).} 
\tablenotetext{c}{Radial component only. Value not corrected for the inclination
angle ($i$) of the outflow axis with respect to the plane of the sky. If correction is
to be done, we recommend $i \sim 10\arcdeg$.}
\tablenotetext{d}{Ambient cloud velocity range.}
\end{deluxetable}

\clearpage

\begin{deluxetable}{cccc}
\tablewidth{0pt}
\tablecolumns{4}
\tablecaption{Mass at different velocity ranges in hh215
\label{masshh215}}
\tablehead{
\colhead{Velocity Range\tablenotemark{a}} &
\colhead{Mass} &
\colhead{Momentum\tablenotemark{b}} &
\colhead{Kinetic Energy\tablenotemark{b}} \\
\colhead{[\kms]} &  
\colhead{[M$_{\sun}$]} &
\colhead{[M$_{\sun}$~\kms]} &
\colhead{[$10^{43}$~erg]}
}
\startdata

$-0.15 < v < 0.74$  &  0.01 & 0.03 & 0.06\\
$1.62 < v < 2.06$   &  0.94 & 0.43 & 0.23\\
$3.16 < v < 4.15$   &  1.31 & 1.25 & 1.28\\
$4.15 < v < 6.46$   &  0.27 & 0.30 & 1.51
\enddata  
\tablenotetext{a}{Velocity ranges are the same ranges as the ones in Figure~6.}
\tablenotetext{b}{Radial component only. Value not corrected for the inclination
angle ($i$) of the outflow axis with respect to the plane of the sky. If correction is
to be done, we recommend $i \sim 10\arcdeg$.}
\end{deluxetable}

\clearpage

\begin{deluxetable}{lp{5.5cm}cccc}
\tabletypesize{\footnotesize} 
\tablewidth{0pt}
\tablecaption{
Comparison of molecular outflow associated with HH~315B
and results from entrainment models 
\label{comphh315b}}
\tablecolumns{6}
\tablehead{
\colhead{} &
\colhead{} &
\colhead{} &
\multicolumn{3}{c}{Consistent with model?}\\
\cline{4-6}
\colhead{Parameter} & 
\colhead{Description} & 
\colhead{Figure} & 
\colhead{Turbulent} & 
\colhead{Bow}  & 
\colhead{Wide-angle} \\
\colhead{} & \colhead{} & \colhead{}  &
\colhead{Jet} & \colhead{Shock}  &  \colhead{Wind}
}
\startdata
Morphology & Small bow-like, with wings extending in direction of outflow source
\newline & 3 
& NO & {\bf YES} & NO\\
Temperature & $T_{ex}$ of outflow  $>$ ambient $T_{ex}$, also increase in $T_{ex}$ with velocity \newline &  11
& NO & {\bf YES} & NO\\ 
Velocity & ``Spur-like''---maximum velocity at position of HH~315B  (shock head) \newline & 13
& NO & {\bf YES} & NO\\
Momentum & Maximum at bow head & 14
& NO & {\bf YES} & NO
\enddata
\end{deluxetable}

\clearpage

\begin{deluxetable}{lp{5.5cm}cccc}
\tabletypesize{\footnotesize} 
\tablewidth{0pt}
\tablecaption{
Comparison of molecular outflow associated with
 HH~315C and results from entrainment models   
\label{comphh315c}}
\tablecolumns{6}
\tablehead{
\colhead{} &
\colhead{} &
\colhead{} &
\multicolumn{3}{c}{Consistent with model?}\\
\cline{4-6}
\colhead{Parameter} & 
\colhead{Description} & 
\colhead{Figure} & 
\colhead{Turbulent} & 
\colhead{Bow}  & 
\colhead{Wide-angle} \\
\colhead{} & \colhead{} & \colhead{}  &
\colhead{Jet} & \colhead{Shock}  &  \colhead{Wind}
}
\startdata
Morphology & Large and wide shell- (or bow-) like, with wings extending in direction of outflow source \newline & 3 & 
NO & {\bf YES} & {\bf YES}\\
Temperature &  $T_{ex}$ of outflow $>$ ambient $T_{ex}$, also increase in $T_{ex}$ with velocity
\newline &  11 & 
NO & {\bf YES} & NO\\
Velocity & Double-peaked velocity distribution \newline & 13  & NO & NO & NO\\
Momentum &  Higher in bow (shell)
  wings due to underlying ambient density distribution 
& 14 & NO & {\bf YES} & {\bf YES}
\enddata
\end{deluxetable}

\clearpage

\begin{deluxetable}{lp{5.5cm}cccc}
\tabletypesize{\footnotesize} 
\tablewidth{0pt}
\tablecaption{
Comparison of redshifted molecular outflow lobe south of PV Ceph
 and results from entrainment models
\label{comphh215red}}
\tablecolumns{6}
\tablehead{
\colhead{} &
\colhead{} &
\colhead{} &
\multicolumn{3}{c}{Consistent with model?}\\
\cline{4-6}
\colhead{Parameter} & 
\colhead{Description} & 
\colhead{Figure} & 
\colhead{Turbulent} & 
\colhead{Bow}  & 
\colhead{Wide-angle} \\
\colhead{} & \colhead{} & \colhead{}  &
\colhead{Jet} & \colhead{Shock}  &  \colhead{Wind}
}
\startdata
Morphology & Jet-like, with wiggling axis \newline & 
   17 & {\bf YES} & {\bf YES} & NO\\
Temperature & Outflow $T_{ex} >$ ambient $T_{ex}$, with local maxima at source position 
and $\sim60\arcsec$ \/ south of source along axis \newline
  & 12  & NO & {\bf YES}\tablenotemark{a} & NO\\
Velocity & Velocity peak at source and  $\sim60\arcsec$ \/ south of source \newline &  15 & NO & {\bf
YES}\tablenotemark{b} & NO\\ 
Momentum & Maximum at source, general decrease with distance from source & 16 & NO & {\bf YES}\tablenotemark{c} & NO \enddata
\tablenotetext{a}{Consistent with bow shock entrainment by several internal working surfaces, each responsible
for the rise in temperature.}
\tablenotetext{b}{Consistent with bow shock entrainment by several internal working surfaces, each responsible
for the rise in velocity. Note, most of the low-velocity redshifted emission at the source position is {\it not}
due to the PV Ceph outflow.}
 \tablenotetext{c}{Consistent with bow shock entrainment with an underlying
ambient density gradient of $r^{-a}$, where $1\leq a \leq 2$.} 
\end{deluxetable}


\clearpage

\begin{thebibliography}{}
\bibitem[Arce \& Goodman 2001a]{ag01a}Arce, H.~G., \& Goodman, A.~A. 2001a,
\apj, 551, L171 
\bibitem[Arce \& Goodman 2001b]{ag01b}Arce, H.~G., \& Goodman, A.~A. 2001b,
\apj, 554, 132
\bibitem[Arce \& Goodman 2002]{ag02}Arce, H.~G., \& Goodman, A.~A. 2002, in press [Paper I]
\bibitem[Bachiller 1996]{b96}Bachiller, R. 1996, \araa, 34, 111
\bibitem[Bachiller \& Tafalla 1999]{bt99}Bachiller, R., \& Tafalla, M.\ 1999, in The Origin of Stars and Planetary Systems,
    ed. N.~D.~Kylafis \& C.~J.~Lada (Dordrecht:Kluwer), 227
\bibitem[Bachiller et al. 1994]{btc94}Bachiller, R., Tafalla, M., \& Cernicharo, J. 1994, \apj, 425, L93
\bibitem[Bence et al. 1996]{brp96} Bence, S.~J., Richer, J.~S., \& Padman, R. 1996, \mnras, 279, 866
\bibitem[Cernicharo \& Reipurth 1996]{cr96} Cernicharo, J., \& Reipurth, B. 1996, \apj, 460, L57
\bibitem[Cabrit \& Raga 2000]{cr97} Cabrit, S., \& Raga, A. 2000, \aap, 354, 667
\bibitem[Cabrit et al. 1997]{crg97} Cabrit, S., Raga, A., \& Gueth, F. 1997, IAU Symp No. 182,
Herbig-Haro Flows and the Birth of Low Mass Stars, ed. B. Reipurth, \& C. Bertout (Dordrecht: Kluwer), 163
\bibitem[Cant\'o \& Raga]{cr91}Cant\'o, J., \& Raga, A.~C. 1991, \apj, 372, 646
\bibitem[Cant\'o et al. 2000]{crd00}Cant\'o, J., Raga, A.~C., \& D'Alessio, P. 2000, \mnras, 313, 656
\bibitem[Chernin \& Masson 1995]{cm95}Chernin, L.~M., \& Masson, C.~R. 1995, \apj, 455, 182
\bibitem[Cliffe et al. 1996]{cfj96}Cliffe, J.~A., Frank, A., \& Jones, T.~W. 1996, \mnras, 282, 1114
\bibitem[Davis et al. 2000]{ddmcm00}Davis, C.~J., Dent, W.~R.~F., Matthews, H.~E., Coulson, I.~M., \& McCaughrean,  M.~J.
2000, \mnras, 318, 952
\bibitem[Davis et al. 1997]{derj97}Davis, C.~J., Eisl\"{o}ffel, J., Ray, T.~P., \& Jenness, T. 1997, \aap, 324, 1013
\bibitem[Devine et al. 1997]{detal97}Devine, D., Bally, J., Reipurth, B., \& Heathcote, S. 1997, \aj, 114, 2095
\bibitem[Davis et al. 1998]{dsm98}Davis, C.~J., Smith, M.~D., Moriarty-Schieven, G.~H. 1998, \mnras, 299, 825
\bibitem[De Young 1986]{d86}De Young, D.~S. 1986, \apj, 307, 62
\bibitem[Downes \& Ray 1999]{dr99}Downes, T.~P., \& Ray, T.~P. 1999, \aap, 345, 977
\bibitem[Fuente et al. 1998]{fetal98}Fuente, A., Mart\'{\i}n-Pintado, J., Bachiller, R., Neri, R., \& Palla,
F. 1998, \aap, 334, 253
\bibitem[Gledhill et al. 1987]{gws87}Gledhill, T.~M., Warren-Smith, R.~F., \& Scarrott, S.~M. 1987, \mnras, 229, 643
\bibitem[G\'omez et al. 1997]{gkd97} G\'omez, M., Kenyon, S., \& Whitney, B.~A. 1997, \aj, 114, 265
\bibitem[Goodman \& Arce 2001]{ga01}Goodman, A.~A., \& Arce, H.~G. 2002, in preparation
\bibitem[Gueth \& Guilloteau 1999]{gg99}Gueth, F., \& Guilloteau, S. 1999, \aap, 343, 571 
\bibitem[Hartigan et al. 2000]{hbrm00}Hartigan, P., Bally, J., Reipurth, B., \& Morse, J.~A. 2000,
in Protostars and Planets IV, ed. V.~Mannings, A.~P.~Boss, \& S.~S.~Russell 
  (Tucson: University of Arizona Press), 867
\bibitem[Hatchell et al. 1999]{hfl99}Hatchell, J., Fuller, G.~A., Ladd, E.~F. 1999, \aap, 344, 687
\bibitem[Heathcote et al. 1996]{hmhrsbs96}Heathcote, S., Morse, J.~A., Hartigan, P., Reipurth, B., Schwartz, R.~D.,
Bally, J., \& Stone, J.~M. 1996, \aj, 112, 1141
\bibitem[Hirose et al. 1991]{husm97}Hirose, S., Uchida, Y., Shibata, K., Matsumoto, R. 1997, \pasj, 49, 193
\bibitem[Kudoh \& Shibata 1995]{ks95}Kudoh, T., \& Shibata, K. 1995, \apj, 452, L41
\bibitem[Kutner \& Ulich 1981]{ku81}Kutner, M.~L., \& Ulich, B.~L. 1981, \apj, 250, 341
\bibitem[Lada \& Fich 1996]{lf96}Lada, C.~J., \& Fich, M. 1996, \apj, 459, 638
\bibitem[Langer et al. 1996]{lvx96}Langer, W.~D., Velusamy, T., \& Xie, T. 1996, \apj, 468, L41
\bibitem[Lee et al. 2000]{lmros00}Lee, C.-F., Mundy, L.~M., Reipurth, B., Ostriker, E.~C., \& Stone, J.~M. 2000, \apj,
542, 925
\bibitem[Lee et al. 2001]{lsom01}Lee, C.~F., Stone, J.~M., Ostriker, E.~C., Mundy, L.~G. 2001, \apj, 557, 429
\bibitem[Leinert et al. 1997]{lrh97}Leinert, C., Richichi, A., \& Hass, M. 1997, \aap, 318, 472
\bibitem[Lery et al. 1999]{lhn99}Lery, T., Heyvaerts, J., Appl, S., Norman, C.~A. 1999, \aap, 347, 1055
\bibitem[Levreault \& Opal 1987]{lo87}Levreault, R.~M., \& Opal, C.~B. 1987, \aj, 93, 669
\bibitem[Li \& Shu 1996]{ls96} Li, Z.-Y., \& Shu, F.~H. 1996, 472, 211
\bibitem[Lizano \& Giovanardi 1995]{lg95}Lizano, S., \& Giovanardi, C. 1995, \apj, 447, 742
\bibitem[Masson \& Chernin 1993]{mc93} Masson, C.~R., \& Chernin, L.~M. 1993, \apj, 414, 230
\bibitem[Matzner \& McKee 199]{mm99} Matzner, C.~D., \& McKee, C.~F. 1999, \apj , 526, L109
\bibitem[Najita \& Shu 1994]{ns94}Najita, J.~R., \& Shu, F.~H. 1994, \apj, 429, 808
\bibitem[Neckel et al. 1987]{nssb87}Neckel, T., Staude, H.~J., Sarcander, M., \& Birkle, K. 1987, \aap, 175, 231
\bibitem[Ostriker et al. 2001]{olsm01}Ostriker, E.~C., Lee, C.-F., Stone, J.~M., Mundy, L.~G. 2001, \apj, 557, 443
\bibitem[Papaloizou \& Terquem 1995]{pt95}Papaloizou, J.~C.~B., \& Terquem, C. 1995, \mnras, 274,
987 
\bibitem[Raga \& Cabrit 1993]{rc93}Raga, A., \& Cabrit, S. 1993, \aap, 278, 267
\bibitem[Raga et al. 1990]{rcbc90}Raga, A.~C., Cant\'o, J., Binette, L., \& Calvet, N. 1990, \apj, 364, 601
\bibitem[Raga \& Kofman 1992]{rk92}Raga, A.~C., \& Kofman, L. 1992, \apj, 386, 222
\bibitem[Reipurth 2000]{r00}Reipurth, B. 2000, \aj, 120, 3177
\bibitem[Reipurth \& Bally 2001]{rb01} Reipurth, B., \& Bally, J. 2001, \araa, 39, 403
\bibitem[Reipurth et al. 1997a]{rbd97} Reipurth, B., Bally, J., \& Devine, D. 1997a, \aj, 114, 2708
\bibitem[Reipurth et al. 1986]{retal86}Reipurth, B., Bally, J., Graham, J.~A., Lane, A.~P., \& Zealey,
W.~J. 1986, \aap, 164, 51 
\bibitem[Reipurth et al. 1997b]{retal97}Reipurth, B., Hartigan, P., Heathcote, S., Morse, J.~A., \& Bally,
J. 1997b, \aj, 114, 757
\bibitem[Reipurth et al. 1992]{rrh92}Reipurth, B., Raga, A.~C., \& Heathcote, S. 1992, \apj, 392, 145
\bibitem[Reipurth et al. 2000]{ryhbr00}Reipurth, B., Yu, K.~C., Heathcote, S., Bally, J., \& Rodr\'{\i}guez, L.~F. 2000, \aj,
120, 1449
\bibitem[Richer et al. 1992]{rhp92}Richer, J.~S., Hills, R.~E., \& Padman, R. 1992, \mnras, 254, 525
\bibitem[Rohlfs \& Wilson 2000]{rw00} Rohlfs, K., \& Wilson, T.~L. 2000, Tools of Radio Astronomy (3rd ed.; New York:
Springer)
\bibitem[Shang et al. 1998]{ssg98}Shang, H., Shu, F.~H., \& Glassgold, A.~E. 1998, \apj, 493, L91
\bibitem[Shu et al. 1991]{srll91}Shu, F.~H., Ruden, S.~P., Lada, C.~J., \& Lizano, S. 1991, \apj, 370, L31
\bibitem[Shu et al. 1995]{snos95} Shu, F.~H., Najita, J., Ostriker, E.~C., \& Shang, H. 1995, \apj, 455, L155
\bibitem[Smith et al. 1997]{ssy97} Smith, M.~D., Suttner, G., \& Yorke, H.~W. 1997, \aap, 323, 223
\bibitem[Stone \& Norman 1993a]{sn93a}Stone, J.~M., \& Norman, M.~L. 1993a, \apj, 413, 198
\bibitem[Stone \& Norman 1993b]{sn93b}Stone, J.~M., \& Norman, M.~L. 1993b, \apj, 413, 210
\bibitem[Stone \& Norman 1994]{sn94}Stone, J.~M., \& Norman, M.~L. 1994, \apj, 420, 237
\bibitem[Suttner et al. 1997]{ssyz97}Suttner, G., Smith, M.~D., Yorke, H.~W., Zinnecker, H. 1997, \aap, 318, 595
\bibitem[Taylor \& Williams 1996]{tw96} Taylor, S.~D., Williams, D.~A. 1996, \mnras, 282, 1343 
\bibitem[Terquem et al. 1999]{tepn}Terquem, C., Eisl\"offel, J., Papaploizou, J.~C.~B., \& Nelson, R.~P. 1999, \apj, 512, L131
\bibitem[Velusamy \& Langer 1998]{vl98}Velusamy, T., \& Langer, W.~D., \nat, 392, 685
\bibitem[Wild 1999]{w99}Wild, W. 1999, A Handbook for the IRAM 30m Telescope, Instituto de Radioastronom\'{\i}a
Milim\'etrica
\bibitem[Wilken 1996]{w96}Wilken, F.~P. 1996, \apj, 459, L31
\bibitem[Wolfire \& K\"onigl 1993]{wk93} Wolfire, M.~G., \& K\"onigl, A. 1993, \apj, 415, 204 
\bibitem[Wu et al. 1996]{whh96} Wu, Y., Huang, M., \& He, J. 1996, 
  A\&AS, 115, 283
\bibitem[Yu et al. 1999]{ybb99}Yu, K.~C., Billawala, Y., \& Bally, J. 1999, \aj, 118, 2940
\bibitem[Zhang \& Zheng 1997]{zz97} Zhang, Q., \&  Zheng, X. 1997, \apj, 474, 719
\end{thebibliography}
\end{document}